\documentclass[onecolumn,authoryear]{els-mrw} 

\usepackage{amsmath,amssymb,amsfonts,amsthm,makeidx,graphicx}
\usepackage{txfonts}
\usepackage{helvet}
\usepackage{xcolor}
\newcommand{\om}{\Omega_{\rm M}}

\newcommand{\oll}{\Omega_{\rm \Lambda}}

\newcommand{\kmsMpc}{{\rm km\,s}^{-1}{\rm Mpc}^{-1}}

\begin{document}

\chapter{Encyclopedia of Astrophysics: The Expanding Universe}\label{chap1}

\author[1]{Tamara Davis}%
%\author[2]{Second Author}%
%\author[1,2]{Third Author}%

\address[1]{\orgname{The University of Queensland}, \orgdiv{School of Mathematics and Physics}, \orgaddress{Brisbane, QLD, 4101 Australia}}
%\address[2]{\orgname{Name of Institute}, \orgdiv{Division or Department}, \orgaddress{Address of Institute}}

\articletag{Encyclopedia of Astrophysics (2026), Volume 5, Chapter 1, pp1-16; 1st Edition; Editor Ilya Mandel; Section editor Cullan Howlett}

\maketitle

\begin{glossary}[Glossary]
\term{Cosmological Principle} The principle that we do not occupy a special place in the universe, and that it is homogeneous and isotropic.\\
\term{Critical Density} Density at which the universe would be flat. \\
\term{Friedmann Equations} Equations governing how the expansion rate of the universe changes with time. \\
\term{Hubble-Lema\^{i}tre Law} Equation for the expansion rate of the universe (recession velocity) being proportional to distance.\\
\term{Peculiar velocity} Any velocity that deviates from the smooth expansion of the Universe.\\
\term{Recession velocity} Velocity of comoving objects moving with the expansion of the Universe.
\end{glossary}

\begin{glossary}[Nomenclature]
\begin{tabular}{@{}lp{34pc}@{}}
BAO &Baryon Acoustic Oscillations\\
CMB &Cosmic Microwave Background\\
FLRW &Friedmann-Lema\^{i}tre-Robertson-Walker, referring to the metric for an homogeneous isotropic space.\\
$\Lambda$CDM &Our standard cosmological model, with a cosmological constant ($\Lambda$) and cold dark matter.

%EPM &Long expression that needs more than one line\hfill\break to complete.\\
\end{tabular}
\end{glossary}

\begin{BoxTypeA}[box:keypoints]{Key Points}
\begin{itemize}
    \item Recession velocity is proportional to distance.   This is known as the Hubble-Lema\^{I}tre Law, $v=HD$.  It is a requirement of the cosmological principle.  Recession velocities can exceed the speed of light without violating relativity, because they are not velocities in any inertial frame. 
    \item The cosmological principle states that on average the Universe is homogeneous and isotropic. It is supported by observations in our Universe.
    \item The Hubble parameter is defined as the expansion rate of the universe divided by the scale factor, $H(a)\equiv\dot{a}/a$.  Hubble’s constant, $H_0$, is its value at the present day ($a=1$);  the current expansion rate of the Universe.  
     %Velocity as a function of redshift is independent of Hubble’s constant. 
    \item Photons are redshifted as they travel through expanding space.  Energy is conserved at every infinitesimal point, but the photons are being measured relative to comoving observers that are  receding. 
    \item Friedmann’s equation describes the dynamics (acceleration) of the expansion.  It can be derived from general relativity, but also from Newtonian equations because most of the Universe is very dilute and therefore in the weak-field regime of gravity; some researchers dispute this. 
    \item The dynamics of the Universe depend on what it is made of, such as matter, radiation, and dark energy.  Different components have different equations of state, $w=p/\rho$, which determines how their energy density evolves, $\rho=\rho_0a^{-3(1+w)}$, and thus their impact on the acceleration.  
    \item The age of the Universe, time of photon travel, distances, and horizons can all be calculated by integrating Friedmann’s equation. 
\end{itemize}
\end{BoxTypeA}
\begin{abstract}[Abstract]
The expansion of the Universe is the basis of modern cosmology.  This chapter outlines the theory behind the expansion of the universe, including the cosmological principle, distances, velocities, and accelerations.  We provide basic derivations of the key equations and highlight some interesting features, such as superluminal expansion, how pressure increases gravitational attraction, the subtleties of conservation of energy in the expanding universe, and the existence of cosmological horizons. 
\end{abstract}

%%%%%%%%%%%%%%%% INTRO %%%%%%%%%%%%%%%%%%%%%
\section{Introduction to the expansion}\label{chap1:sec1}

The discovery of the expansion of the Universe in the 1920s was the birth of the modern era of cosmology.  It followed hot on the heels of the discovery that the `spiral nebulae' were not clouds of spiraling gas nearby, but entire stellar systems far away -- now known as galaxies.  Several researchers including Vesto Slipher, Henrietta Swan Leavitt, and Milton Humason helped generate the data used by Georges \citet{Lemaitre1927} and Edwin \citet{Hubble1929} to discover the expansion.  They used redshifts to observe that most galaxies are receding from us, and that the more distant the galaxies the faster they are receding.  That relationship has now become known as the Hubble-Lema\^{i}tre Law,
\begin{align}
    v=HD, 
\end{align}
where $v$ is the recession velocity, $H$ is the Hubble parameter, and $D$ is the proper distance.  The Hubble parameter has dimensions of inverse time but is usually quoted as the velocity of expansion in km\,s$^{-1}$ at the distance of 1~Mpc (i.e.~$\kmsMpc$). If the expansion neither accelerates nor decelerates then it is easy to use the Hubble-Lema\^{i}tre Law to calculate how much time has passed since the beginning of the Universe, $v=D_0/t_0=H_0D_0$ so the age of the Universe is $t_0=1/H_0$ (subscript 0 indicates the present day).  This remains a good order-of-magnitude approximation even in the presence of moderate accelerations; we derive the age of the Universe in general in Sec.~\ref{sec:age}.

The Hubble-Lema\^{i}tre Law is required for an homogeneous isotropic expansion.  If velocity was {\em not} proportional to distance, then an homogeneous distribution would become inhomogeneous with time. This can be derived from the general metric for an homogeneous isotropic space, which is known as the Friedmann-Lema\^{i}tre-Robertson-Walker (FLRW) Metric, 
\begin{align}\label{eq:FLRW}
    ds^2 = -c^2 dt^2 + R(t) \left[d\chi^2 + S_k(\chi)d\psi^2\right],
\end{align}
where $ds$ in the space-time interval, $c$ is the speed of light, $dt$ is the time that a comoving observer observes on their watch, $R$ is the scalefactor, $d\chi$ represents the radial comoving coordinate and $d\psi^2 \equiv d\theta^2 + \sin^2\theta d\phi^2$ represents the perpendicular comoving coordinates.  The factor $S_k(\chi)\equiv \sin(\chi), \chi, \sinh(\chi)$ for positive, flat, and negative curvature universes respectively. 

Comoving observers are an important concept in this framework.  They are observers who are moving {\em only} due to the expansion of the Universe and do not have any other velocity. The comoving coordinates are labels to positions in space, which get further or closer to each other depending on the evolution of the scalefactor.  Comoving observers are those that stay at constant comoving coordinates, that is their only velocities are recession velocities due to the expansion of space.  We introduce the word `dust' as a technical term to indicate an homogeneous distribution of comoving matter particles. 

In Equation~\eqref{eq:FLRW} the scalefactor $R$ has dimensions of distance.  There are many ways to write the FLRW metric, and I have summarized many of them in Box~\ref{box:FLRW}.  A common variant uses the normalised scalefactor $a=R/R_0$, so at the present day $a(t_0)\equiv a_0=1$.

The Hubble-Lema\^{i}tre Law can be derived from Equation~\eqref{eq:FLRW} by setting $dt=0$ and $d\psi=0$.  This leaves the radial distance along a constant time slice, and integrating gives, 
\begin{align}\label{eq:dist}
    D=R\chi.
\end{align}
Differentiating Eq.~\ref{eq:dist} gives $v=\dot{R}\chi$. Equating this to $v = HD$ defines the Hubble parameter: $H\equiv\dot{R}/R=\dot{a}/a$.  Thus the Hubble-Lema\^{i}tre Law is a simple consequence of the symmetries of expanding homogeneous, isotropic space.

%---------------------------------
%%%%%%%%%%%%%%%% BOX FLRW %%%%%%%%%%%%%%%%%%%%%
\begin{BoxTypeA}[box:FLRW]{The many forms of the Friedmann-Lema\^{i}tre-Robertson-Walker Metric}

The Friedmann-Lema\^{i}tre-Robertson-Walker (FLRW) metric comes in several equivalent forms.  The main differences between them being the definition of comoving distance, and which variable carries the dimensions.  Seven main variations are described below with a description of the links between them.  The list is not comprehensive, but includes the most common and most useful forms.
\begin{align}
ds^2 = -c^2dt^2 + R^2(t)\left[d\chi^2 + S_k^2(\chi) d\psi^2\right] \qquad
^{\mbox{$\chi=\frac{c}{R_0}\int\frac{dz}{H(z)}$ - {\em dimensionless}}}_{\mbox{$R$ - dimensions of {\em distance}}}
\label{xR}
\end{align}

Set $r=S_k(\chi)$.  
Differentiating gives $d\chi^2=\frac{dr^2}{1-kr^2}$, where $k=-1,0,1$ for open, flat, closed respectively.  %(On a unit sphere, $x^2+y^2=1$, i.e.\ when $r=\sin(\chi)$ $dr/d\chi = \cos(\chi)= \sqrt(1-r^2)$.  For a hyperboloid, $x^2-y^2=1$, etc...)
\begin{align} 
ds^2 = - c^2dt^2 + R^2(t)\left[\frac{dr^2}{1-kr^2} + r^2 d\psi^2\right] \qquad
^{\mbox{$r$ - {\em dimensionless}}}_{\mbox{$R$ - dimensions of {\em distance}}}
\label{rR}
\end{align}

Set $R(t)=R_0a(t)$ and $x=R_0\chi$.  Substitute into Equation~\ref{xR}.
\begin{align} 
ds^2 = -c^2dt^2 + a^2(t)\left[dx^2 + R_0^2S_k^2\left(\frac{x}{R_0}\right) d\psi^2\right] \qquad
^{\mbox{$x$ - dimensions of {\em distance}}}_{\mbox{$a$ - {\em dimensionless}}}
\label{chia}
\end{align}

Set $R(t)=R_0a(t)$ and $\mbox{\sf{r}}=R_0r$.  Substitute into Equation~\ref{rR}.
\begin{align} 
ds^2 = -c^2dt^2 + a^2(t)\left[\frac{d{\sf r}^2}{1-k{\sf r}^2/R_0^2} + {\sf r}^2 d\psi^2\right] \qquad
^{\mbox{$\sf{r}$ - dimensions of {\em distance}}}_{\mbox{$a$ - {\em dimensionless}}}
\label{sfrR}
\end{align}

Define $\kappa = k/R_0^2$, which is the curvature parameter that can take on any value, so closed, flat, and open universes correspond to $\kappa>0, =0, <0$ respectively.
\begin{align} 
ds^2 = -c^2dt^2 + a^2(t)\left[\frac{d{\sf r}^2}{1-\kappa{\sf r}^2} + {\sf r}^2 d\psi^2\right] \qquad
^{\mbox{$\sf{r}$ - dimensions of {\em distance}}}_{\mbox{$a$ - {\em dimensionless}}}
\label{eq:FLRWkappa}
\end{align}

Set $d\eta=cdt/R(t)$.  Substitute into Equation~\ref{xR}.
\begin{align} 
ds^2 = R^2(t)\left[-d\eta^2 + d\chi^2 + S_k^2(\chi) d\psi^2\right] \qquad
^{\mbox{$\chi$ and $\eta$ - {\em dimensionless}}}_{\mbox{$R$ - dimensions of {\em distance}}}
\label{etaR}
\end{align}

Set $dn=cdt/a(t)$.  Substitute into Equation~\ref{chia}.
\begin{align} 
ds^2 = a^2(t)\left[-dn^2 + dx^2 + R_0^2S_k^2\left(\frac{x}{R_0}\right) d\psi^2\right] \qquad
^{\mbox{$x$ and $n$ - dimensions of {\em distance}}}_{\mbox{$a$ - {\em dimensionless}}}
\label{na}
\end{align}
Different situations are more easily addressed with different metrics.  For example, when calculating radial distance, then $x$ or $\chi$ (\ref{xR},\ref{chia}) would be simpler to deal with than $r$ or ${\sf r}$ (\ref{rR},\ref{sfrR}) because there are no curvature components in the radial part of the metric.  On the other hand, the calculation of angular size distance and luminosity distance which depend on the angular components of the metric would be simplified by using $r$ or ${\sf r}$.
It is often easier to deal with the present day scalefactor as normalised to 1 (i.e.\ $a$), and attribute dimensions to the comoving coordinates.
Changing the time coordinate to conformal time $\eta$ or $n$ (\eqweblink{Equations}{etaR} and \eqref{na}) includes time in the expansion, so past light cones have a gradient of 1 and comoving worldlines are straight.  Note that you cannot arbitrarily set $R_0=1$ {\em and} $k=-1,0,1$.  The curvature parameter $\kappa=k/R_0^2$ is negative, zero, or positive in open, flat and closed universes respectively. 
\end{BoxTypeA}

\begin{figure}%[t]
\centering
\includegraphics[width=1.0\textwidth]{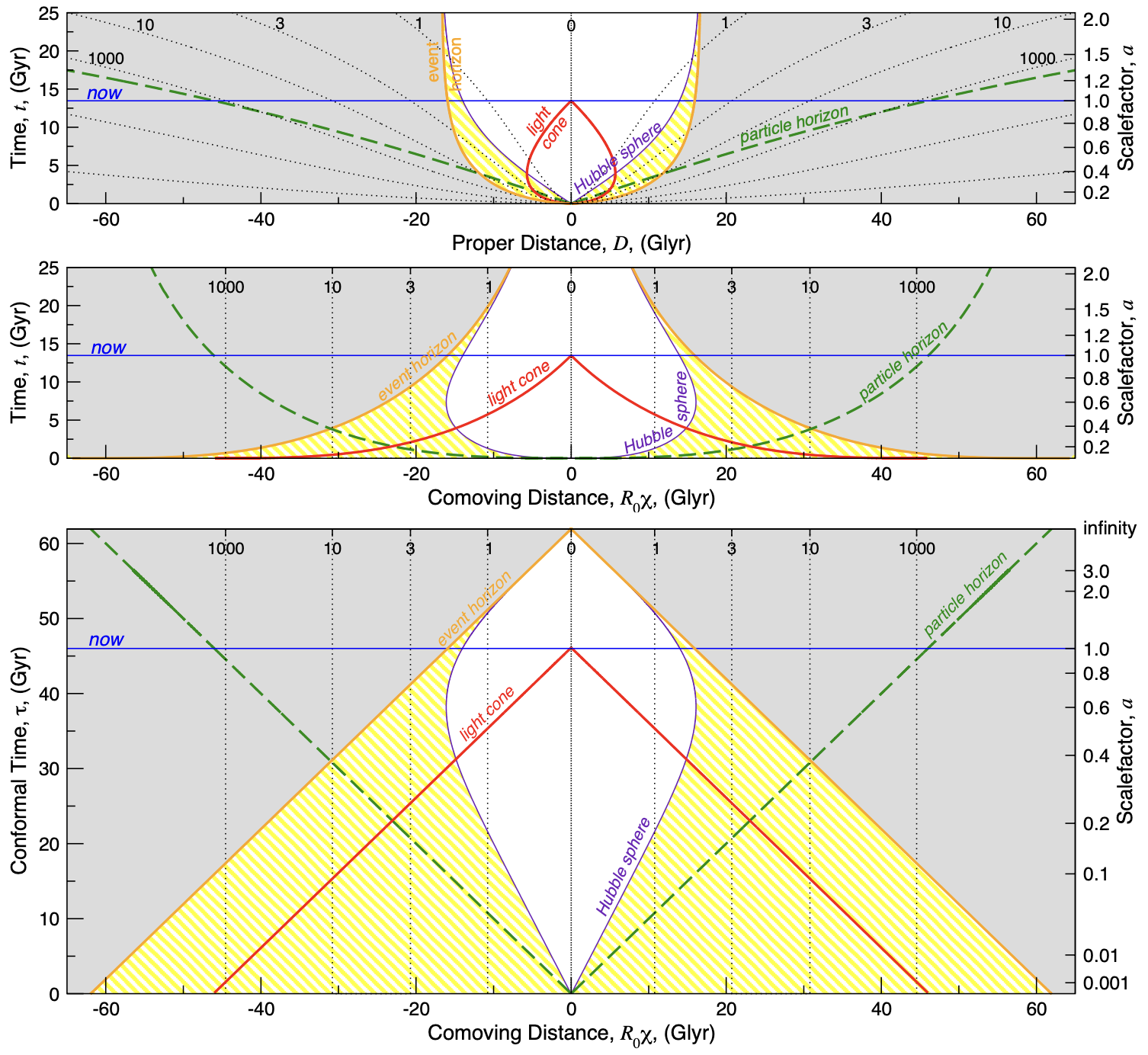}
\caption{Spacetime diagrams for the $(\om,\oll) = (0.3,0.7)$ universe with $H_0 = 70~\kmsMpc$. Dotted lines show the worldlines of comoving objects. The current redshifts of the comoving galaxies appear labeled at the top of each comoving worldline. The normalized scalefactor, $a = R/R_0$, is drawn as an alternate vertical axis. {\em Our} comoving coordinate is the central vertical worldline. All events that we currently observe are on our past light cone (the cone or ``teardrop'' with apex at $t = $now). All comoving objects beyond the Hubble sphere (thin solid line) are receding faster than the speed of light. The speed of photons on our past light cone relative to us (the slope of the light cone) is not constant, but is rather $v-c$. Photons we receive that were emitted by objects beyond the Hubble sphere were initially receding from us (outward sloping lightcone at $t \lesssim 5$~Gyr, upper panel). Only when they passed from the region of superluminal recession $v > c$ (yellow crosshatching and beyond) to the region of subluminal recession (no shading) could the photons approach us. More detail about early times and the horizons is visible in comoving coordinates (middle panel) and conformal coordinates (lower panel). Our past light cone in comoving coordinates appears to approach the horizontal ($t = 0$) axis asymptotically, however it is clear in the lower panel that the past light cone reaches only a finite distance at $t = 0$ (about 46~Glyr, the current distance to the particle horizon). Light that has been traveling since the beginning of the universe was emitted from comoving positions which are now 46~Glyr from us. The distance to the particle horizon as a function of time is represented by the dashed green line. This is the distance to the most distant object we are able to observe at any particular time. Our event horizon is our past light cone at the end of time, $t = \infty$ in this case. It asymptotically approaches $\chi = 0$ as $t \rightarrow \infty$. Many of the events beyond our event horizon (shaded solid gray) occur on galaxies we can see (the galaxies are within our particle horizon). We see them by light they emitted billions of years ago but we will never see those galaxies as they are today. Galaxies with redshift $z \sim 1.8$ are just now passing over our event horizon. Galaxies with redshift $z \sim 1.45$ are now receding at the speed of light. The vertical axis of the lower panel shows conformal time (proper time divided by the scalefactor). 
An infinite proper time is transformed into a finite conformal time so this diagram is complete on the vertical axis. 
The aspect ratio of $\sim 3/1$ in the top two panels represents the ratio between the size of the universe and the age of the universe, which in this model is 46~Glyr/13.5~Glyr.}
\label{chap1:SpacetimeDiagrams}
\end{figure}

\begin{BoxTypeA}[box:Ht]{The `present day expansion rate' decreases during acceleration!}

The Hubble parameter $H(t)$ changes with time; the Hubble constant $H_0$ is its value at the present day (it is a constant at all positions in space, but not constant in time).  %In other words, at any snapshot in time the Hubble parameter takes a single value at all positions in space, but at different times $H_0$ would take different values.
$H_0$ is often described as the {\em current} expansion rate of the Universe, since $a=1$ at the present day, so $H_0=\dot{a}$.  This might make you think that if the Universe accelerates, $H(t)$ would increase with time.  However, one has to remember that the denominator ($a$) increases too, and it typically increases more quickly than the numerator ($\dot{a}$), unless the expansion is increasing at a {\em higher} than exponential rate.  So in our Universe $H(t)$ decreases with time. The Hubble parameter is a constant in {\em time} only if the universe is expanding at an exactly exponential rate.  (This can be seen if you set $a=e^{H_0t}$, so $\dot{a}/a=H_0$.)  

For a visual explanation of decreasing $H(t)$, see the spacetime diagrams in Fig.~\ref{chap1:SpacetimeDiagrams}. The dotted lines are the worldlines of comoving galaxies. Imagine drawing a vertical line at $D=20$~Glyr in the top panel. As time goes on the velocity at which comoving galaxies cross the $D=20$~Glyr line decreases (the slopes become more vertical).  Thus, the recession velocities at a constant distance decrease.
\end{BoxTypeA}

%---------------------------------
%%%%%%%%%%%%%%%% PHILOSOPHY %%%%%%%%%%%%%%%%%%%%%
\section{Philosophical basis}
The basis of modern cosmology is the Cosmological Principle --- the idea that humans do not hold a privileged position in the Universe. It is an extension of the Copernican revolution, which demoted Earth from a special position at the centre of the solar system and instead declared that all planets orbit the Sun.  Of course we now know that this is an approximation, and they actually orbit their common center of mass.  

The Cosmological Principle holds true if the Universe is homogeneous and isotropic. Homogeneous means it is the same in all positions, and isotropic means it is the same in all directions.  Of course, this too is only an approximation, yet it seems to be a good one on large enough scales.  If you average over large volumes of space (larger than a sphere of radius about 100~Mpc), then the Universe does indeed appear to be homogeneous and isotropic \citep{Scrimgeour2012,Ntelis2017}.  In my cosmology lectures I often joke that the basis of modern cosmology is the assumption that we do not exist, after all, we are extreme deviations from homogeneity.  However, when calculating the dynamics of the expansion of the Universe, assuming homogeneity and isotropy results in simple calculations that appear to do a good job of explaining how the expansion rate of the Universe changes over time.  The dynamics are governed by the Friedmann equations, which we will derive in  Section~\ref{sec:friedmann}. 

There are papers that dispute this claim \citep[e.g.][]{Wiltshire2007,Buchert2008,Giani2024}. %Aluri2023
They say that the inhomogeneities are important and that non-linear impacts of general relativity cannot be neglected when considering large-scale dynamics.  This kind of effect is known as `backreaction'.  However, the prevailing opinion is that the backreaction effects should be small \citep[e.g.][]{Ishibashi2006,Baumann2012,Kaiser2017}.  One of the reasons for this is that the Universe is very dilute; the average density at the present day is only a few atoms per square metre.  That means most gravitational fields are very weak.  Thus the dynamics of the expansion can be calculated in the weak-field regime of relativity --- namely Newtonian gravity (see Section~\ref{sec:newtonian}).  

Philosophically, the Cosmological Principle is very satisfying.  It can be reworded to state that our Universe is (on average) the same at every position and in every direction.  However, we could go one step further and expect our Universe to follow the {\em perfect} cosmological principle, which would require the Universe to also be the same at all {\em times}.  The `Steady State Theory' is a cosmological model that obeys the perfect cosmological principle by creating matter spontaneously in the vacuum of space as the Universe expands. 

Observationally we see that our Universe falls short of obeying the perfect cosmological principle.  Instead we find ourselves in a Universe that is expanding and that {\em had a beginning}.\footnote{By this we do not necessarily mean there was nothing before the Big Bang, as our theories do not yet extend to describe the universe before about $t=10^{-43}s$ (the Planck time), but rather that the Universe was once in an extremely hot, dense state from which it expanded.}       
Nevertheless, for decades after the expansion was discovered the `Steady State Theory' was very popular.  This was in part because the value of the Hubble constant $H_0$ had been (erroneously) estimated to be several hundred $\kmsMpc$, which would make the age of the Universe only a couple of billion years old -- younger than the geological time records on Earth!  The Steady State Theory allowed there to be expansion without a beginning, which solved that problem.  However, when better measurements of $H_0$ showed a Big Bang Universe was conceivably older than the stars it contains, the Big Bang Theory started to gain in popularity.  In the Big Bang Theory the Universe began from a hot dense state and expanded.  It was dramatically confirmed by the discovery of the Cosmic Microwave Background (CMB).  The CMB is the relic radiation left over from that hot dense early Universe, and its discovery in the 1960s heralded the death of the Steady State Theory. 

The age of the universe remained a problem until the 1990s, however, because a universe that decelerated to its current expansion rate would only have been 9 or 10 billion years old, whereas observations of stars and globular clusters indicates that they are older than that.  That was solved by the discovery of the acceleration of the universe in the late 1990s \citep{Riess1998,Perlmutter1999}, because a universe that accelerates to its current rate of expansion is older than one that purely decelerated (see Fig.~\ref{chap1:fig:at}). We give the name `dark energy' to whatever is causing the accelerated expansion. The leading dark energy candidate is the cosmological constant ($\Lambda$), which could be due to vacuum energy \citep{Carroll1992}.  Dark energy also solved another problem. Theorists had noted that if the universe is close to flat it should be exactly flat, because flatness is an unstable point in a decelerating universe. Observers could only find enough matter and dark matter to make up about 30\% of the energy density needed for flat universe, the inclusion of dark energy supplies the other 70\%.  Our standard cosmological model is thus an homogeneous, isotropic expansion governed by general relativity, and dominated by cold dark matter along with a cosmological constant, $\Lambda$CDM.

%%%%%%%%%%%%%%%% REDSHIFT %%%%%%%%%%%%%%%%%%%%%
\section{Redshift}
The observation that galaxies are receding is the observation that their light is redshifted.  Redshift is defined as,
\begin{align}
    z=\frac{\lambda_0 - \lambda_{e}}{\lambda_{e}} = \frac{\lambda_0}{\lambda_{e}}-1
\end{align}
where $\lambda_0$ is the wavelength at the present day and $\lambda_e$ is the wavelength at the time of emission.  Wavelengths stretch in proportion to scalefactor,\footnote{To see this take the equation for comoving distance (Eq.~\ref{eq:distcomovt}) and consider a source emitting photons at a constant intervals $\delta t_{\rm e}$ that are observed at intervals $\delta_{\rm t_o}$.  The comoving distance remains the same, so 
\begin{align}
R_0\chi=c\int_{t_{\rm e}}^{t_{\rm o}}\frac{dt}{a(t)} =c\int_{t_{\rm e}+\delta t_{\rm e}}^{t_{\rm o}+\delta t_{\rm o}}\frac{dt}{a(t)}.
\end{align}
As long as the intervals are small the integrals differ by only, $\frac{\delta t_{\rm o}}{a(t_{\rm o})}-\frac{\delta t_{\rm e}}{a(t_{\rm e})}$, which has to be zero.  Thus, $\frac{\delta t_{\rm o}}{\delta t_{\rm e}}=\frac{a(t_{\rm o})}{a(t_{\rm e})} = \frac{\lambda_{\rm o}}{\lambda_{\rm e}},$
where the last term comes from the fact that frequency is inversely proportional to time and wavelength is inversely proportional to frequency.  This derivation follows that of \citet{Hartle2003} (Sec.~18.2); for a derivation from general relativity see \citet{Carroll2004} (Sec.~3.5).
} % TODO:{\red [cite derivation, or derive]},
$\lambda\propto a$, so redshift is directly related to the relative size of the Universe at the time of emission compared to the time of absorption,
\begin{align}\label{eq:redshift}
    z=\frac{1}{a}-1
\end{align}
where $a$ is the scalefactor at the time of emission.  Thus the redshift of a photon that was emitted when the universe was half of its present size ($a=1/2$) is $z=1$, whereas light that was emitted at the beginning of the universe ($a=0$) would have $z=\infty$.  That means the amount of expansion (change in scalefactor) between $0<z<1$ is the same as the amount of expansion between $1<z<\infty$.  The distance to the most distant object we can see is called the particle horizon (see Sec.~\ref{sec:horizon}).  While in theory that corresponds to light emitted at the beginning of the universe, and thus $z=\infty$, in practice we cannot see that far because the early universe was opaque.  The earliest we can see is the CMB, which is at $z\sim1090$.  If we were able to observe primoridal neutrinos we would be able to see back to earlier times, because the universe was transparent to neutrinos from about a second after the big bang. 

In the presence of other sources of redshift, such as from peculiar velocities ($z_{\rm p}$), one finds that the observed redshift ($z_{\rm o}$) is related to the cosmological redshift ($z$) by,\footnote{To derive this, consider a comoving observer at the position of the source, who observes the source to have wavelength $\lambda_{\rm c}$.  Then the redshift observed can be split into the redshift the comoving observer sees, followed by the redshift due to the expansion: $1+z_{\rm o}=\frac{\lambda_{\rm o}}{\lambda_{\rm e}} =\frac{\lambda_{\rm o}}{\lambda_{\rm c}} \frac{\lambda_{\rm c}}{\lambda_{\rm e}}=(1+z)(1+z_{\rm p})$.} 
\begin{align}\label{eq:zcombo}
    (1+z_{\rm o})= (1+z)(1+z_{\rm p}).
\end{align}

%%%%%%%%%%%%%%%% AGE %%%%%%%%%%%%%%%%%%%%%
\section{Calculating age}\label{sec:age}
\begin{figure}[h]
\centering
\includegraphics[width=.6\textwidth]{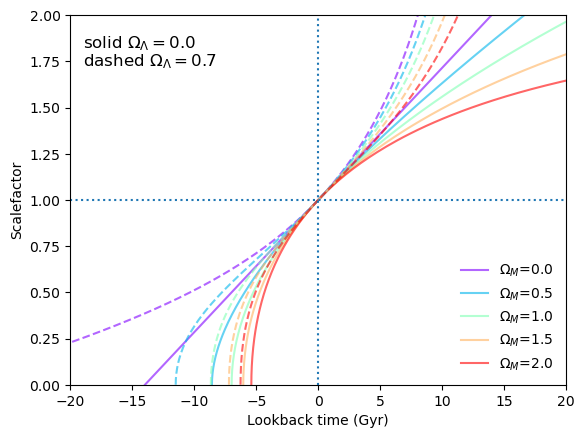}
\caption{How scalefactor evolves with time in various universes with matter density as shown in the legend and cosmological constant either 0.0 (solid lines) or 0.7 (dashed lines).  Densities are defined in Sec.~\ref{sec:norm}. The horizontal axis is lookback time in billions of years, and all universes have been normalised to have the same expansion rate of 70~$\kmsMpc$ at the present day.  The age of the universe can been seen by when the scalefactor hits zero.  Universes with higher matter density are younger than those with lower matter density, because they decelerate more strongly.  Adding a cosmological constant increases the age of the universe, because of the period of acceleration before the present day.}
\label{chap1:fig:at}
\end{figure}

The age of the Universe can be calculated once we know how the Hubble parameter changes with scalefactor (see Section~\ref{sec:friedmann}).  Working from the definition of $H$ we see, 
\begin{align}
    \frac{da}{dt} \equiv Ha, \\
    \int_{t_1}^{t_2} dt = \int_{a_1}^{a_2} \frac{da}{Ha}.
\end{align}
To calculate the age of the Universe as a function of scalefactor set the lower limits of the integrals to zero, and the upper limits to $t$ and $a$.  To calculate the age at the present day set the upper limits to the current age and current scalefactor (i.e. $t_2=t_0$ \& $a_2=1$).  To calculate the lookback time, which is how long ago something happened, use the current age and scalefactor in the upper limits, but set the lower limits to the scalefactor and time you want to calculate.  Similarly, if you want to know how long ago something  at redshift $z$ occurred, calculate the time between $a_1=1/(1+z)$ and $a_2=1$.
A plot of scalefactor vs lookback time is shown in Fig.~\ref{chap1:fig:at} for various universes.

%%%%%%%%%%%%%%%% DISTANCE %%%%%%%%%%%%%%%%%%%%%
\section{Calculating distance}
To figure out the distance along a path that a photon has followed, one has to set $ds=0$ in the metric.  Looking first at purely radial distance (setting $d\psi=0$) one finds,
\begin{align}
    0=-cdt + R(t)d\chi,\\
    \int_{\chi_1}^{\chi_2} d\chi%= c\int_{R_1}^{R_2}\frac{dt}{R(t)} 
    = \frac{c}{R_0}\int_{t_1}^{t_2}\frac{dt}{a(t)}, \label{eq:distcomovt}\\
    R_0\Delta\chi = c\int_{a_1}^{a_2} \frac{da}{a\dot{a}}.\label{eq:R0chi}
\end{align}
It is often useful to work in terms of redshift.  From Eq.~\ref{eq:redshift} we know $a=1/(1+z)$ so,
\begin{align}
   \frac{da}{dz} = -(1+z)^{-2}. 
\end{align}
We can thus replace the integral over scalefactor in Eq.~\ref{eq:R0chi} with $da=-(1+z)^{-2}dz$ and we get the distance {\em at the present day} of an object seen at redshift $z$, known as the comoving distance,
\begin{align}\label{eq:distcomov}
   D(z,t_0)=R_0\chi(z) = c\int_{0}^{z} \frac{dz}{H(z)},
\end{align}
where we have swapped the order of the integral to remove the minus sign and set the final redshift to its present day value $z_2=0$ (putting $a=1$ into Eq.~\ref{eq:redshift}). We drop the $\Delta$ for conciseness since we can arbitrarily set the start of the path as $\chi_1=0$.  Note the redshift in the upper limit of the integral should be the redshift purely due to the expansion of the universe (as opposed to the observed redshift, which can be contaminated with contributions from peculiar velocities). 
Eq.~\ref{eq:distcomov} has dimensions of distance so is known as the comoving distance (as opposed to $\chi$ which is the comoving coordinate).  

The proper distance at any time is,
\begin{align}\label{eq:distprop}
    D(z,t)=R(t)\chi(z)=a(t)R_0\chi(z).
\end{align}  There is a unique comoving distance for every redshift. The scalefactor gives the scaling needed to calculate the distance at any other time.  For example, to calculate the distance at the time the light we now see was emitted, use $a(t) = 1/(1+z)$.

\subsection{Luminosity and Angular Diameter Distance}\label{sec:lumang}

There are a couple of other `distances' that are in common use.  These are not true distances that appear in any metric, but they are designed to make some of our Euclidean expectations for luminosity and size work, and are related to observables. 

{\bf Luminosity Distance} ($D_{\rm L}$) is defined such that the observed flux (energy per second per area squared) is equal to the emitted luminosity (energy per second) divided by the surface area of a sphere with radius luminosity distance.  In other words, it makes the inverse square law work: Flux = Luminosity$/(4\pi D_{\rm L}^2)$.  Contrary to expectations in a non-expanding universe, the energy of photons decreases not only because of spreading out over the surface of a sphere, but also because they get redshifted and time-dilated.  Each of those contributes a factor of  $(1+z_o)$ (where $z_o$ is the {\em observed} redshift), so the luminosity decreases by an additional $(1+z_o)^2$. In addition, due to curvature the surface area of a sphere should use the distance $RS_k(\chi)$ instead of the radial distance $R\chi$.  You can see this by integrating the metric along a constant radius but all angles to get the surface area of the sphere.  Thus Flux = Luminosity$/(4\pi R^2S_k^2(\chi)(1+z_o)^2)$, which means Luminosity Distance at the present day is,
\begin{align}\label{eq:distlum}
D_{\rm L}=R_0S_k(\chi)(1+z_o).
\end{align}

{\bf Angular Diameter Distance} ($D_{\rm A}$) is defined such that the observed angular size of a source is the intrinsic size divided by the angular diameter distance.  In other words, it makes the normal angle formula work: $\theta = x/D_{\rm A}$, where $\theta$ is the angle subtended by the source and $x$ is the size of the source perpendicular to the line of sight.  For the purposes of this calculation we again need to know the perpendicular distance $RS_k(\chi)$, but we need to know it at the time of emission. The size the object appears is determined by its distance when it was emitted, not the distance it is now.  Thus we need to use $R=R_0a(t_{\rm em})=R_0/(1+z_o)$.  The Angular Diameter Distance at the present day is therefore,
\begin{align}\label{eq:distlum}
D_{\rm A}=R_0S_k(\chi)/(1+z_o).
\end{align}

Note that $D_{\rm A}=D_{\rm L}(1+z_o)^2$.  This is known as the Etherington relation \citep{Etherington1933}, or `distance duality'.  It will be true in any metric theory of gravity where photons are conserved, because the difference between angular diameter distances and luminosity distances is simply to swap the role of source and observer. 

The time dilation, redshift, and apparent size effects depend on the {\em observed} redshift. Therefore I have been careful to note the $(1+z_o)$ factors that appear in luminosity distance and angular diameter distance are the observed redshifts.  Due to peculiar velocities and gravitational redshifts the observed redshift may differ from the cosmological redshift that appears in the upper limit of the integral for comoving distance.  In detail, one could make that clearer in the equations by writing $D_{\rm L}(z,z_o)=R_0S_k(\chi(z))(1+z_o)$ and $D_{\rm A}(z,z_o)=R_0S_k(\chi(z))/(1+z_o)$.  

%%%%%%%%%%%%%%%% VELOCITY %%%%%%%%%%%%%%%%%%%%%
\section{Calculating the velocity in the Hubble-Lema\^{i}tre Law}
Multiplying Eq.~\ref{eq:distcomov} by $H_0$ gives the full equation for present day recession velocity that is valid at all redshifts, 
\begin{align}
    v(z)=c\int_0^z\frac{dz}{E(z)},
\end{align}
where $E(z)\equiv H(z)/H_0$ and $H(z)$ is the Hubble parameter as a function of redshift, which we derive in Sec.~\ref{sec:friedmann}.  (When we say ``as a function of redshift'' we are using redshift to refer to the {\em time} that the object at a redshift $z$ emitted the light we now see.)  Note that $E(z)$ is a function of cosmological parameters such as matter density, but is independent of $H_0$.  Thus the recession velocity as a function of redshift is {\em independent of the current expansion rate}, $H_0$. 

In the 1920s when expansion was first measured, the recession velocity could be approximated by $v=cz$ because all the measured galaxies were so close to us that this low-redshift approximation was sufficient.  A better approximation that is reasonably good out to a redshift of about $z\sim0.1$ is,
\begin{align}\label{eq:vzHD}
    v(z) = \frac{c z}{1+z} \left[1+\frac{1}{2}(1-q_0)z - \frac{1}{6}(1-q_0-3q_0^2+j_0)z^2\right], 
\end{align}
with the deceleration parameter $q_0=-0.55$ and jerk parameter $j_0=1.0$ for standard $\Lambda$CDM with matter density and cosmological constant densities $(\Omega_{\rm M},\Omega_{\rm \Lambda})\sim(0.3,0.7)$.  Eq.~\ref{eq:vzHD} is often used for nearby redshifts because it is faster than calculating the full integral.\footnote{Sometimes it is claimed that using Eq.~\ref{eq:vzHD} makes the result independent of the cosmological model.  However, because the values of $q_0$ and $j_0$ need to be tuned to a cosmological model, it  does not actually remove the cosmological model dependence.}  

So far we have covered the kinematics of the Universe such as measures of distance and velocity.  In the next sections we embark on understanding the dynamics of the expansion, that is how it accelerates or decelerates according to the gravitational effects of the universe's contents. We begin by examining how densities of different components of the universe evolve, for which we invoke conservation of energy.

%%%%%%%%%%%%%%%%% DENSITY AND PRESSURE %%%%%%%%%%%%%%%%
\section{How densities and pressures change with the expansion: conservation of energy}\label{sec:conservation}

A critical part of the calculation of the dynamics of the universe is understanding how densities and pressure change as the Universe expands.  This is different for different materials.  For dust (an homogeneous gas of comoving particles) the only energy of relevance is the rest mass of the particles, because the particles are comoving. As the universe expands their energy density drops in proportion to the volume, that is, in proportion to $\rho_{\rm M}\propto a^{-3}$.  For radiation the particles are moving at the speed of light and as the universe expands not only does the number density drop in proportion to volume but also their wavelengths stretch in proportion to the scalefactor.  This means that their energy density drops by an extra factor of scalefactor, $\rho_{\rm R}\propto a^{-4}$. Massive particles that have large peculiar velocities, such as neutrinos, are in between the behaviour of dust and radiation. Meanwhile, dark energy in the form of a cosmological constant does not dilute as the universe expands, so its energy density is $\rho = {\rm constant}$. 

Each of these different forms of energy density affect how the expansion of the universe accelerates.  We can use the first law of thermodynamics (conservation of energy) to calculate how densities and pressures change as the universe expands. The general relativistic result comes from the zeroth component of the equation for conservation of energy in general relativity, 
\begin{align}\label{eq:Tmunu0}
    \nabla_\mu T^{\mu}_0=0
\end{align} 
\citep[see][Eq.~8.52]{Carroll2004}.\footnote{$0=\partial_\mu T^{\mu}_{0}+\Gamma^{\mu}_{\mu\lambda}T^{\lambda}_{0}-\Gamma^{\lambda}_{\mu 0}T^{\mu}_{0\lambda}=-\partial_t\rho-3\frac{\dot{a}}{a}(\rho+p)$.} The same result comes from applying the first law of thermodynamics, which gives the rate of change of energy, $E$, as volume, $V$, changes,
\begin{align}
    \frac{dE}{dt}=-p\frac{dV}{dt}.
\end{align}
We apply this to any sphere in the universe with arbitrary radius $r=R\chi$ and $V=4\pi (R\chi)^3/3$.  Using the fact that energy is equal to energy density\footnote{Remember we are using $\rho$ to be {\em energy} density; if it was matter density then energy density would be $\rho c^2$.} times volume, $E=\rho V$, we can replace $V$ with $R^3$ (because the $4\pi\chi^3/3$ cancels from both sides) to get,
\begin{align}
    \frac{d(\rho R^3)}{dt} = -p\frac{dR^3}{dt}. %-\frac{p}{c^2}\frac{dR^3}{dt}.
\end{align}
Rearranging and using $a=R/R_0$ gives,
\begin{align}\label{eq:rhodot}
\frac{\dot{\rho}}{\rho} = -3(1+w)\frac{\dot{a}}{a},
\end{align}
where the `equation of state' is defined\footnote{If we had not set $c=1$ it would be $w=p/(\rho c^2)$.} as $w=p/\rho$.   Finally, integrating gives, 
\begin{align}
    \rho=\rho_0 a^{-3(1+w)}.
\end{align}
This equation shows how density changes with scalefactor as the universe expands. That change is dependent on the equation of state, which relates density to pressure.  Different components of the Universe have different equations of state.  Radiation, matter, and cosmological constant have respectively $w=1/3,0,-1$.  

The cosmological constant, or `vacuum energy', has a negative equation of state, which means it has a negative pressure.  This is a very odd behaviour, as it means that if you had a piston of vacuum energy and compressed it slightly it would suck the piston in instead of pushing it back out.  In terms of the effect on the universe, it has the odd effect of having repulsive gravity.  

Having established the relationship between density and pressure as the universe expands we now calculate its dynamics. To do so we embark briefly on an adventure into general relativity. 

\begin{BoxTypeA}[box:pressure]{The Effect of Pressure on Acceleration}

Usually when we consider pressure we think of high pressure pushing outwards.  Therefore we might expect a high pressure to accelerate expansion. However, that intuition only applies when there is a {\em pressure gradient}.  The high pressure needs somewhere to go.  High pressure gas will push a piston out into lower pressure gas, until the pressures are equal.  However, in the homogeneous universe there are no pressure gradients.  There is nowhere for high pressure gas to go.  So pressure does not do any work pushing outwards.  

Instead, high pressure {\em pulls inwards} due to gravity.  This is because all forms of energy gravitate.  If something has pressure it means the particles in it are moving around.  The kinetic energy of their motion is an extra source of gravity in addition to their rest mass.  Therefore, higher pressure corresponds to an increased gravitational attraction of the gas. 

This applies to more than just gas.  Radiation has an equation of state of $w=1/3$, and therefore it is more effective at {\em decelerating} the universe compared to dust, which has $w=0$.  Dark energy, on the other hand, has an equation of state of $w=-1$, which means it has negative pressure.  Since positive pressure gravitates, negative pressure anti-gravitates.  Thus dark energy has a {\em repulsive} gravitational effect, and {\em accelerates} the expansion of the universe. 
\end{BoxTypeA}

%%%%%%%%%%%%%%%% FRIEDMANN %%%%%%%%%%%%%%%%%%%%%
\section{The Friedmann Equations}\label{sec:friedmann}
The Friedmann equations describe how the expansion rate of the Universe changes with time due to the gravitational effects of the contents of the universe.  In any universe that is not empty, the expansion will typically accelerate or decelerate.  The full derivation of the dynamics of the expansion comes from general relativity, but an intuitive derivation can also be made using simple Newtonian physics.  Since the Universe is mostly very dilute,  the weak-field limit of relativity applies in most cases.  Here we will outline both derivations and explore the implications for the expanding Universe.

%%%%%%%%%%%%%%%% FRIEDMANN GR %%%%%%%%%%%%%%%%%%%%%
\subsection{General relativitstic derivation of Friedmann's first Equation}

In this section we outline the general relativistic derivation of the first Friedmann equation (often simply called the Friedmann equation).  We assume basic familiarity with General Relativity and its usual notation; for a more complete derivation see, for example \citet{Carroll2004,Moore2013}.
The dynamics of the Universe are governed by the Einstein Field Equation,
\begin{align} 
G_{\mu\nu} = 8\pi T_{\mu\nu}.
\label{eq:einstein}
\end{align}
Where the Einstein tensor can be broken into, 
\begin{align}
    G_{\mu\nu} = R_{\mu\nu}-\frac{1}{2} g_{\mu\nu} R -\Lambda g_{\mu\nu}.
    \label{Eq:Gmunu}
\end{align}
We use the stress energy momentum tensor of a perfect fluid to describe the contents of the Universe, $T_{\mu\nu}=(\rho+p)u_\mu u_\nu - pg_{\mu\nu}$, where $u_\mu$ is the 4-velocity or,
\begin{align}
T_{\mu\nu} = 
\begin{pmatrix}
\rho c^2 & 0 & 0 & 0 \\
0 & p & 0 & 0 \\
0 & 0 & p & 0 \\
0 & 0 & 0 & p
\end{pmatrix},
\label{eq:Tmunu}
\end{align}
and we will use the FLRW metric in the form of Eq.~\ref{eq:FLRWkappa}.  We set $c=1$ and use $\rho$ to mean the energy density.

Using the symmetries of the equation we can calculate the non-zero components of the Ricci tensor,
\begin{align}
    R_{tt} &= -3\frac{\ddot{a}}{a},\\
    R_{rr} &= \frac{a\ddot{a}+2\dot{a}^2+2\kappa}{1-\kappa r^2}, \label{eq:RicciR}
         \\
    R_{\theta\theta} &= r^2(a\ddot{a}+2\dot{a}^2+2\kappa), \\
    R_{\psi\psi} &= r^2(a\ddot{a}+2\dot{a}^2+2\kappa)\sin^2\theta,
    \label{eq:RicciT}
\end{align}
and the Ricci scalar is,
\begin{align}
    R = 6\left(\frac{\ddot{a}}{a}+\frac{\dot{a}^2}{a^2}+\frac{\kappa}{a^2}\right).
    \label{eq:RicciS}
\end{align}
The first Friedmann equation comes from the temporal part of the Einstein Equation,
\begin{align}
    R_{tt}-\frac{1}{2}Rg_{tt}-\Lambda g_{tt} = 8\pi G T_{tt}.
\end{align}
A factor of $3\ddot{a}/a$ cancels because it appears with opposite signs in the first two terms.  The remaining terms are,
\begin{align}
    3\left(\frac{\dot{a}}{a}\right)^2 +3\frac{\kappa^2}{a^2}-\Lambda = 8\pi G\rho,
    \end{align}
which rearranges to become the common form of the first Friedmann equation, showing how the Hubble parameter changes with scalefactor,
\begin{align}
   H^2= \left(\frac{\dot{a}}{a}\right)^2 = \frac{8\pi G}{3}\rho+\frac{\Lambda}{3}-\frac{\kappa^2}{a^2}.
   \label{eq:Friedmann1gr}
\end{align}

%%%%%%%%%%%% FRIEDMANN NEWTONIAN %%%%%%%%%%%%%%
\subsection{Newtonian derivation of the first Friedmann equation}\label{sec:newtonian}

In this section we will use conservation of energy and the continuity equation to derive Friedmann's equation.  Since Newtonian physics is the weak-field limit of General Relativity, and the universe is quite dilute, the Newtonian derivation exactly matches the result from general relativity. 
We will perform the derivation for non-relativistic matter to build intuition, but it is easily generalised to other contents of the universe. 

Start with conservation of energy for a test particle of mass $m$ and velocity $v$ at a radius $D$ from the centre of an arbitrary sphere of total mass $M$ in the homogeneous, isotropic space. The sum of kinetic and potential energy is,
\begin{align}\label{eq:etot}
    E_{\rm total}=\frac{1}{2}mv^2-\frac{GMm}{D}. 
\end{align}
Note that according to Birkoff's theorem the matter inside a sphere can be considered a point source at the origin for the purposes of calculating its gravitational effect. 
The mass within an homogeneous sphere of density $\rho$ is given by $M=\frac{4}{3}\pi D^3 \rho$.  Moving the constants to the left of the equation, 
\begin{align}
    \frac{2 E_{\rm total}}{m}=v^2 - \frac{8\pi G D^2}{3}\rho.
\end{align}
Then setting $v=\dot{R}\chi$ and $D=R\chi$ (and again pulling constants to the left) gives,
\begin{align}
    \frac{2 E_{\rm total}}{m\chi^2}=\dot{R}^2 - \frac{8\pi G R^2}{3}\rho.
\end{align}
The left hand side of that equation can be absorbed into a constant, which we will call $-k$ (chosen to match the general relativistic derivation). %$k$ is related to the curvature of the universe. 
Dividing by $R^2$ gives Friedmann's equation,
\begin{align}\label{eq:Friedmann1}
    H^2 = \frac{8\pi G}{3}\rho -\frac{k}{R^2} \\= \frac{8\pi G}{3}\rho -\frac{\kappa}{a^2} .
\end{align}
This equation is the same as the general relativistic derivation, except that we left out the cosmological constant. To generalise this  to also include a cosmological constant one simply uses $\rho = \rho_{\rm M}a^{-3}+\rho_\Lambda$. In general, to add radiation or any other component (like dark energy with $w\ne1$) one uses $\rho=\sum_x \rho_x a^{-3(1+w_x)}$; see Sec.~\ref{sec:timevarying} for what to do if $w$ changes with time.

Some conclusions about the fate of the expansion of the universe can be made from this derivation, in the special case when the universe is exclusively made up of non-relativistic matter.  

In a `dust'-filled universe when $E_{\rm total}>0$, the kinetic energy is greater than potential energy and the expansion will continue forever.  For $E_{\rm total}<0$ the expansion will eventually stop, turn around, and recollapse. The case where $E_{\rm total}=0$ is perfectly balanced and will asymptote to no expansion at infinite time. 

While the correspondence with the fate of the Universe is {\em only true for dust-only universes} (adding cosmological constant, dark energy, or radiation break the perfect correspondence), the curvature conclusions are general.  
$E_{\rm total}=0,>0,<0$ corresponds to $\kappa=0,<0,>0$, which are spatially flat, negatively curved, and positively curved cases respectively.  By scaling the value of $R_0$ one can tune $\kappa$ such that it becomes exactly $k=R_0^2\kappa=0,-1,+1$.

\subsection{Critical density and normalised densities}\label{sec:norm}
The case where $E_{\rm total}=0$ corresponds to a flat universe with $k=0$.  This is a special case, and we define what is known as the {\em critical density}, $\rho_{\rm c}$, as the density required to make the universe flat.  From Eq.~\ref{eq:Friedmann1} with $k=0$,
\begin{align}
    \rho_{\rm c}=\frac{3H^2}{8\pi G}.
\end{align}
It is convenient to define densities relative to this critical density at the present day, namely,
\begin{align}
    \Omega_x = \frac{\rho_{x0}}{\rho_{\rm c0}}.
\end{align}
In these units a flat universe has a total density of $\Omega_{\rm T}=1$, which can be made up of various components.  In our standard cosmological model with a cosmological constant and cold dark matter, $\Lambda$CDM, we find the matter density\footnote{Matter density includes both dark matter and normal `baryonic' matter, the latter has $\Omega_{\rm B}\sim0.05$.} $\Omega_{\rm M}\sim0.3$, the dark energy density in the form of a cosmological constant to be $\Omega_{\Lambda}\sim0.7$, and a trace of radiation with $\Omega_{\rm R}\sim 8\times 10^{-5}$.  Interestingly, even though the energy density of photons is so much lower than that of matter, the number of photons still exceeds the number of protons in the universe by a factor of a billion.\footnote{Wien's displacement law, tells us the peak wavelength of a blackbody occurs at $\lambda_{\rm peak}=0.002898/T$~Km, which gives the typical energy per photon. Since photon energy$=hc/\lambda$ and the CMB is approximately 2.725~K, the energy per photon is approximately $2\times10^{-22}$~J.  Whereas that of a proton is $m_p c^2\sim1.5\times 10^{-10}$~J.  You can use the ratio of $\Omega_{\rm B}$ to $\Omega_{\rm R}$ to see that the total energy density in baryons is smaller than that difference.  So even if the number of photons from stars was negligible, the number of photons from the CMB vastly outnumbers the number of protons in the universe.  This is also related to the matter-antimatter asymmetry, as matter-antimatter annihilation produced many photons in the early universe.}

\begin{figure}[h]
\centering
\includegraphics[width=.5\textwidth]{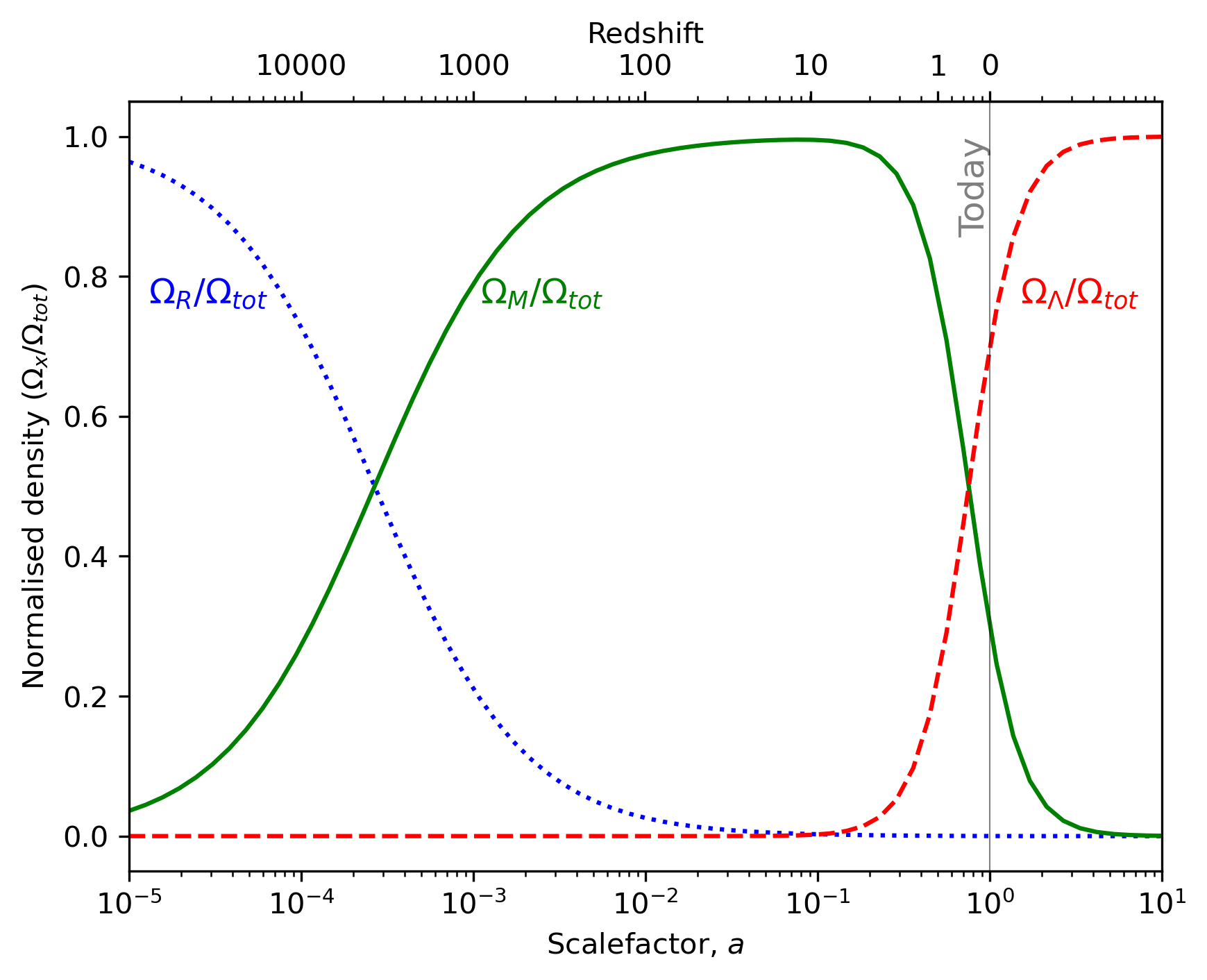}
\caption{This shows how the density of radiation (blue dotted line), matter (green line), and cosmological constant (red dashed line) evolve relative to the total density in a Flat-$\Lambda$CDM model with the standard values of $(\Omega_{\rm R},\Omega_{\rm M}, \Omega_{\Lambda})=(8\times10^{-5},0.3,0.7)$.  Even though radiation is negligible at the present day it dominated before a redshift of about 3300. A commonly cited conundrum in our cosmological model is the `coincidence problem'.  We find ourselves in one of the brief periods of the universe where matter density and cosmological constant density are of similar importance.  This is often considered strange because the vast majority of the time in the universe we are dominated by just one constituent.  Some dark energy and dark matter models attempt to explain this coincidence by invoking (for example) interacting dark components or structure evolution. }
\label{chap1:fig:density}
\end{figure}

In general both the critical density and the densities of each component change with time.  When referring to the time-variable densities (rather than the present day values) we will use $\Omega_x(t)$.  These are shown for our standard cosmological model in Fig.~\ref{chap1:fig:density}, in which I show $\Omega_x(t)/\Omega_c(t)$.

\newpage
\subsubsection{Converting Friedmann's equations to normalised densities (critical density)}
Here we show how to convert the Friedmann equation to normalised densities, which are often in practice more convenient to deal with than plain energy density.  Start from Eq.~\ref{eq:Friedmann1gr} 
% The Friedmann Equation is given by,
%\begin{align}
%\left(\frac{\dot R}{R}\right)^2=\frac{8\pi G\rho}{3}+\frac{\Lambda}{3}-\frac \kappa{R^2} .\label{eq:fried} 
%\end{align}
and substitute $\rho_{\rm M}=\rho_{\rm M 0}a^{-3}$ for matter.\footnote{For calculations of the early universe it is important to include radiation as well, adding the term $\rho_{\rm R}=\rho_{\rm R0}a^{-4}$, but we will leave this out for simplicity.}  
To match the units of the other densities we can define the energy density of the cosmological constant to be $\rho_\Lambda = \Lambda/(8\pi G)$.
\begin{align}
H^2-\frac{8\pi G\rho_{\rm M0}a^{-3}}{3}-\frac{8\pi G \rho_{\Lambda 0}}{3}&=\frac{-k}{R_0^2a^2}. 
\end{align}
Then use the critical denstiy $\rho_{\rm c0}=3H_0^2/(8\pi G)$, to convert that to a normalised density $\Omega = \rho_0/\rho_{\rm c0}$. 
%R_0^2a^2H_0^2\left[\frac{H^2}{H_0^2}-\frac{8\pi G\rho_o}{3H_0^2}a^{-3}-\frac{\Lambda}{3}\right]&=&-kc^2 \\
\begin{align}
a^2H_0^2\left[\frac{H^2}{H_0^2}-\om a^{-3}-\oll\right]&=-\frac{k}{R_0^2}=-\kappa \label{eq:fried3}.
\end{align}
Note there is still a pesky unnormalised scalefactor $R_0$ in this version of the equation.
Since $\kappa$ is a constant it is convenient to evaluate it at the present epoch, 
$ H_0^2\left[1-\om -\oll\right] = -\kappa$ and substitute it back into (\ref{eq:fried3}).
%\begin{align} 
%R_0^2\;a^2H_0^2\left[\frac{H^2}{H_0^2}-\om a^{-3}-\oll\right] = R_0^2H_0^2\left[\Omega_{\rm M}+\Omega_{\Lambda}-1\right].
%\end{align}
Conveniently, this leaves an equation that uses neither the unnormalised scalefactor $R$ {\em nor} the curvature parameter $\kappa$, (recall $H\equiv\dot{a}/a$),
\begin{align}
\dot{a}=H_0\left[1+\Omega_{\rm M}\left(\frac{1}{a}-1\right)+\Omega_{\Lambda}(a^2-1)\right]^{1/2}.
\end{align}
Since a flat universe is one in which the sum of all the $\Omega_x$ components equals 1, the curvature is sometimes quantified as another density parameter $\Omega_{\rm K}$ such that $\Omega_{\rm M}+\Omega_\Lambda+(\Omega_{\rm anything\, else}) + \Omega_{\rm K} = 1$.   So, in the absence of anything else, 
\begin{align} \Omega_{\rm K} = 1-\Omega_{\rm M}-\Omega_\Lambda. \end{align}
This lets us write Friedmann's equation as, $\dot{a}=H_0\left[\Omega_{\rm K}+\Omega_{\rm M}\frac{1}{a}+\Omega_{\Lambda}a^2\right]^{1/2}.$
Using the equation of state $w\equiv p/\rho$ means that for general fluids you can write $\Omega_x = \Omega_{x0}a^{-3(1+w)}$.  This allows us to write this in the most convenient condensed form,
\begin{align} 
H^2 = H_0^2 \sum\Omega_x a^{-3(1+w)}, \label{eq:Ha}
\end{align}
where the way we have defined $\Omega_{\rm K}$ means `curvature' has an effective equation of state, $w_{\rm K}=-1/3$.
%The component of the universe that I have left out of Eq.~\ref{eq:frieddadt} is important to include for early-time calculations is radiation. 

%%%%%%%% TIME VARYING %%%%%%%%%%%%%
\subsubsection{Time varying equations of state}\label{sec:timevarying}
The derivation above applies only for constant equations of state.  Neutrinos have a time-varying equation of state moving from $w\sim1/3$ to $w\sim0$ as they move from relativistic to non-relativistic during the expansion.  Dark energy may also have a time-varying equation of state.  Friedmann's equation for a varying $w$ is given by Eq.~\ref{eq:Ha} with the following replacement,
\begin{equation} a^{-3(1+w)} \rightarrow \exp \left(3\int_a^1\frac{1+w(a)}{a}da\right) \quad \equiv \quad \exp \left(3\int_0^z[1+w(z)]d \ln (1+z)\right).\label{eq:aarrow} \end{equation}
Using the popular parameterization for dark energy $w(a)=w_0+w_a(1-a)$ \citep{Chevallier2001,Linder2003}, Eq.~\ref{eq:aarrow} simplifies to,
\begin{align} a^{-3(1+w_0)} \rightarrow a^{-3(1+w_0+w_a)}e^{-3w_a(1-a)}. \end{align}

\subsection{The second Friedmann Equation}

The first Friedmann equation gave the expansion rate as a function of scalefactor, $\dot{a}$.  The second Friedmann equation gives acceleration $\ddot{a}$.
It can be derived from any spatial part of the metric, because they all generate the same equation.  Choosing the radial direction, 
\begin{align}
R_{rr}-\frac{1}{2}Rg_{rr}-\Lambda g_{rr} = 8\pi G T_{rr},
\end{align}
where $T_{rr}=\rho g_{rr}$, and it is useful to reformulate the radial component of the Ricci tensor (Eq.~\ref{eq:RicciR}) as $R_{rr}= \left(\frac{\ddot{a}}{a}+2\frac{\dot{a}^2}{a^2}+2\frac{\kappa}{a^2}\right)g_{rr}$.  
%\begin{align}
%\left(\frac{\ddot{a}}{a}+2\frac{\dot{a}^2}{a^2}+2\frac{\kappa}{a^2}\right)g_{rr}-\frac{1}{2}6\left(\frac{\ddot{a}}{a}+\frac{\dot{a}^2}{a^2}+\frac{\kappa}{a^2}\right)g_{rr}-\Lambda g_{rr}  = 8\pi G p g_{rr}.\\
%\end{align}
Canceling the $g_{rr}$ and collecting terms leaves,
\begin{align}
    \frac{\ddot{a}}{a}+\frac{3\dot{a}^2}{2a^2}+\frac{3\kappa}{2a^2}-\Lambda=4\pi G p.
\end{align}
Using Eq.~\ref{eq:Friedmann1gr} to remove $\frac{\dot{a}}{a}$ gives the second Friedmann equation, showing how acceleration depends on density and pressure,
\begin{align}
    \frac{\ddot{a}}{a} = -\frac{4\pi G}{3}(\rho+3p)+\frac{\Lambda}{3}.
\end{align}
To derive the second Friedmann equation without reference to general relativity, simply note the once you have used conservation of energy to derive $\dot{\rho}$ (Eq.~\ref{eq:rhodot}) the second Friedmann equation is the time derivative of the first!

\begin{BoxTypeA}[box:conservation]{Conservation of energy in the expanding Universe}

Let's look at conservation of energy in the expanding Universe a little more closely.  If energy is to be conserved in the universe as a whole one would expect it to be conserved in every comoving volume (because if energy is lost from one comoving volume it has to be gained by another, and they would all need to be in equilibrium as required by homogeneity).  However, we have seen in Sec.~\ref{sec:conservation} that energy density evolves with scalefactor in different ways for radiation, matter, and dark energy.  To get the total energy in a comoving volume, one just multiplies the energy density by the volume.  Volume is proportional to scalevactor cubed, $V\propto a^{3}$.  Thus the energy of matter, radiation, and cosmological constant in a comoving volume evolve as,
\begin{align}
    \rho_{\rm M}\propto a^{-3} &\Rightarrow E_{\rm M}={\rm constant},\\
    \rho_{\rm R}\propto a^{-4} &\Rightarrow E_{\rm R}\propto a^{-1}, \\
    \rho_{\rm \Lambda}={\rm constant} & \Rightarrow E_{\Lambda}\propto a^{3}.   
\end{align}
Evidently, the energy in `dust'-like matter is conserved in a comoving volume, because although the density dilutes the volume increases by the same amount.  However, the energy in radiation drops, because not only does the number density of photons drop in proportion to the volume, but each photon loses energy as it is redshifted, which contributes another factor of $a$.  On the other hand, the energy contributed by the cosmological constant {\em increases}, because the vacuum does not dilute as the universe expands (thus the name `cosmological constant').  

There is no way these can add up to a constant, so it would seem that conservation of energy is violated!  However, we used conservation of energy to derive the Friedmann equations, so how is this possible?  It arises because the definition of conservation of energy in general relativity only has a differential form.  $\nabla_\mu T^{\mu}_0=0$ states that energy is conserved at every infinitesimal point and in every interaction throughout the universe.  However, it is not possible to integrate over a constant time slice and find a {\em volume} in which energy is conserved.  At heart this is because of the ambiguity of simultaneity in relativity.  It is impossible to define an unambiguous time slice across which energy should be conserved.  Indeed the time coordinate we have chosen in FLRW has some strange properties --- it requires comoving observers to agree on the time since the big bang, and also that they all agree on the density of the universe at all times.  While convenient, this is not the only time choice we could have made.  In relativity one expects time-dilation between moving observers, and since comoving galaxies really are moving away from each other, they should not share the same time.  

This is related to the question of ``Where does the energy go when photons are redshifted by the expansion of the universe?"  The answer is that it does not go anywhere -- their energy is simply being measured by observers who are receding from the source.  Due to the relative velocity of source and emitter the photon's energy appears to get lower, but no energy has been lost.  %cite Davis 2010 Scientific American?

Noether's theorem is the foundation for conservation of energy.  It states that if there exists time symmetry, conservation of energy must hold.  Time symmetry holds in every infinitesimal interaction, which is encapsulated in the $\nabla T_{\mu\nu}=0$ of general relativity.  however, the expansion of the universe violates time symmetry, so that there is no total energy we can define that will be conserved. 
Some special situations have symmetries that allow an unambiguous total energy to be defined in general relativity.  Such spaces have a `timelike Killing vector'.  The universe is not one of them. 
\end{BoxTypeA}

\section{Horizons}\label{sec:horizon}
There is a limit to how far we can see in the expanding Universe. Three types of horizon are worth noting, in particular. 

{\bf Particle horizon}
The particle horizon is the distance light has traveled from the beginning of the universe until now (or until any time $t$).  It can be calculated by taking the limits in Eq.~\ref{eq:R0chi} from $a=0$ to $a=1$. It determines which distances (which comoving particles) we can currently see.  

{\bf Event horizon}
The event horizon is the distance light can travel from now (or any time $t$) until the end of the universe. It can be calculated by taking the limits in Eq.~\ref{eq:R0chi} from $a=1$ to $a=\infty$. It determines which events we will ever be able to see.  It only exists in universes that expand at an accelerating rate -- they must asymptote to at least exponential expansion.  Interestingly galaxies that are inside our particle horizon can be currently beyond our event horizon (see the galaxies in the grey shaded region of Fig.~\ref{chap1:SpacetimeDiagrams} that are also closer to us than the green dashed particle horizon).  We can see them by light they emitted in the past, but we will never see them as they are today.  Our event horizon is at approximately $z\sim1.8$, which means most of the galaxies in images such as the Hubble Deep Field have already crossed the event horizon and we will never see them as they are at the present day.

{\bf Hubble sphere}
The Hubble sphere is the distance at which galaxies are receding at the speed of light.  It exists in any universe, and is defined as $D_H=c/H$. It is not strictly a horizon, because we can see beyond the Hubble sphere unless the universe is accelerating at a greater than exponential rate. That means there are galaxies with $v>c$ that we can observe.  Interestingly in a universe that is expanding at faster than exponential rate there are galaxies with $v<c$ that end up beyond our event horizon. 

\section{The fate of the universe}
The fate of the universe depends on the laws of physics that govern its expansion and the contents of the universe to which those laws apply.  Consistent with the equations outlined in this chapter, it is possible to create universes that `bounce', universes that recollapse, universes that expand forever, and universes that accelerate at an increasing rate.  Observations must inform our understanding of which type of universe we live in.  So far they indicate we live in a universe that after an initial burst of rapid acceleration settled into a decelerating phase that was slowly taken over by the repulsive force of some sort of dark energy.  This transition occurred because initially as matter dominated the universe the attractive force of gravity was winning the tug-of-war, but as matter diluted the feeble push of dark energy began to dominate.  If this model is true we live in a universe that will expand forever at an accelerating rate and settle into a smooth exponential expansion as matter density tends to zero.  This is known as a `big freeze'.  The full cornocopia of possible universe fates is as follows.

{\bf Big Crunch:}  In high matter density universes the expansion will eventually stop, turn around, and recollapse to a reverse big bang, known as a big crunch.  (Sometimes jokingly referred to as the gnab gib, because it is the big bang backwards.)

{\bf Big Freeze:} If the matter density is not high enough to stop the expansion the universe will expand forever, getting ever colder and fizzling out to `heat death' when we have reached the maximum entropy state.  This applies to both decelerating and accelerating universes as long as they are in the sweet spot where they are not dense enough to recollapse and not accelerating fast enough to ever exceed exponential expansion. Interestingly, the maximum entropy state in accelerating universes of this type (which generally asympote to exponential expansion) will {\em not} be homogeneous gas, even after all black holes evaporate.  That is because they have event horizons and causality will prevent the gas from ever homogenising.  Unfortunately that does not mean that we can extract work forever, as there will still be a maximum entropy state.

{\bf Big Bounce:} For high dark energy universes, the acceleration can be so large that to reach our current expansion rate there will never have been a big bang.  These universes began in a collapsing state, reached a minimum size, and then started expanding again.  

{\bf Big Rip:} This unlikely case describes universes that end up with faster than exponential expansion.  In such cases the event horizon does not asymptote at a constant distance, as is the expectation for our universe, but instead find the event horizon asymptoting to zero radius.  Since nothing can be larger than the event horizon, that means that all structures, including potentially atoms, would get ripped apart.  This is an unlikely conclusion because universes that have faster than exponential expansion violate some important energy principles and have `ghosts', which make them theoretically unstable.  

{\bf Big Wiggle:} Some lesser-known models predict an oscillating Universe. These might have non-standard physics that causes a big crunch to bounce, or have a finely tuned parameters that result in oscillations continuing forever (e.g. $-0.5<\Omega_\Lambda <0, w>0$, and high matter density). 

Observations will reveal which type of universe we live in.  Those without a big bang are already ruled out by the existence of the CMB.  The details of our fate depend on the properties of dark energy and dark matter, which are yet to be sufficiently constrained.  However, the current standard model predicts we will end in a Big Freeze with exponential expansion. Interestingly, if humans emerged later on in our universe, it might be difficult for them to discover that we live in an expanding universe at all.  The CMB will have been redshifted into un-observability, and all the receding galaxies will have disappeared beyond our event horizon (the light we continue to receive from them from the past also redshifted into un-observability).  The only evidence of the outside universe will be our local group galaxies to which we are gravitaionally bound.  Without observing redshifts or the CMB it is hard to imagine how those future humans would deduce the existence of expansion and the origin of the universe in a big bang.  That should also serve as a caution --- what aspects of the universe are {\em we} missing because of our specific position in time in this evolving universe?

\section{Conclusions}
This chapter has outlined the main derivations of the mathematics of our expanding universe.  There is much left to learn.  At the time of writing there are several observational `tensions' that indicate that our model of a homogeneous isotropic universe dominated by cold dark matter and a cosmological constant may need revision \citep[for a reveiw, see][]{Valentino2021}.  

The most prominent tension is that the expansion rate of the universe, $H_0$, as measured by nearby objects on the distance ladder %{\red [cite chapter?]} ($H_0\sim73\kmsMpc$) 
exceeds that estimated from cosmological fits to the cosmic microwave background and baryon acoustic oscillations  %($H_0\sim67-68 \kmsMpc$) 
by about $5\sigma$ \citep{Riess2022,Planck18_VI}. % {\red [cite chapter?]}.  
This could instead be a tension between standard candles and standard rulers, as most of the local measurements are candles while the distant ones are rulers. By distance duality (Sec.~\ref{sec:lumang}) luminosity and angular diameter distances differ by a factor of $(1+z)^2$.  So if there were a bias in our redshifts (for example due to large scale structure) it would bias the candles and rulers in opposite directions \citep{Davis2019}. 

Other tensions include the `$\sigma_8$' tension ($\sigma_8$ is the dispersion of density measured in $8\,h^{-1}$Mpc spheres), in which measurements of the strength of clustering from gravitational lensing disagree with those from galaxy clustering \citep{Hikage2019,DES2020,BOSS2021,KIDS2021} --- although this may have been resolved by the latest measurements from \citet[][]{KIDS2025}.  There is also the `bulk flow' or `dipole' tensions in which peculiar velocities seem larger than expected \citep{Whitford2023,Watkins2023}.  Finally, we have seen hints both in standard candles \citep{des2024} and standard rulers \citep{desi2024} that dark energy may not be a cosmological constant, but might vary in time.  Enormous new data sets being gathered over the next few years will enable us to make ever more precise and accurate measurements of the properties of the expansion. That will reveal whether these tensions are real and we need an improved cosmological model or whether they were results of our hubris and underestimation of our systematic errors.

The expansion of the universe is a fascinating subject that literally holds the story of our origin and the fate of our world.  It is sure to throw up many more surprises in the future. 
%\section{Other types of distance}

%\section{Peculiar velocity decay}

%\section{Other things}
%Is the Universe a black hole?

\begin{ack}[Acknowledgments]

This work was performed as part of the Australian Research Council Centre of Excellence for Gravitational Wave Discovery (project number CE230100016) and funded by the Australian Government.  I thank Leonardo Giani and Cullan Howlett for suggestions which improved this manuscript and Caitlin Ross for calculating which universes are oscillating.

\end{ack}

%\seealso{article title article title}
\def \aapr{Astron.~Astrophys.~Rev.}
\def \aaps{Astron.~Astrophys.~Supp.}
\def \aap{Astron.~Astrophys.}
\def \ag{Astron.~Geophys.}
\def \aj{Astron.~J.}
\def \al{Astron.~Lett.}
\def \anp{Ann.~Phys.}
\def \apss{Astrophys.~Space~Sci.}
\def \apb{Appl.~Phys.~B}
\def \apjl{Astrophys.~J.}
\def \apjs{Astrophys.~J.~Supp.}
\def \apj{Astrophys.~J.}
\def \ap{Appl.~Phys.}
\def \araa{Annu.~Rev.~Astron.~Astrophys.}
\def \arns{Annu.~Rev.~Nuc.~Sci.}
\def \ar{Astron.~Rep.}
\def \asp{Astron.~Soc.~Pac.}
\def \baas{Bull.~Am.~Astron.~Soc.}
\def \baps{Bull.~Am.~Phys.~Soc.}
\def \bist{Bull.~Inf.~Sci.~Tech.}
\def \ca{Comments~on~Astrophys.}
\def \cjc{Canadian~J.~Chem.}
\def \cjp{Canadian~J.~Phys.}
\def \cqg{Classical~Quantum~Gravity}
\def \cup{Cambridge~Univ.~Press}
\def \el{Electron.~Lett.}
\def \epjc{Euro.~Phys.~J.~C}
\def \epjd{Euro.~Phys.~J.~D}
\def \epjst{Euro.~Phys.~J.~Special~Topics}
\def \gca{Geochimica~et~Cosmochimica~Acta}
\def \grg{Gen.~Relativ.~Gravitation}
\def \ijmpd{Int.~J.~Mod.~Phys.~D}
\def \jcap{J.~Cosmo.~Part.~Phys.}
\def \jcp{J.~Chem.~Phys.}
\def \jesrp{J.~Electron~Spectrosc.~Related~Phenomena}
\def \jhep{J.~High~Energy~Phys.}
\def \jms{J.~Molecular~Spectrosc.}
\def \josb{J.~Opt.~Soc.~Am.~B}
\def \jos{J.~Opt.~Soc.~Am.}
\def \jpb{J.~Phys.~B}
\def \jpcrds{J.~Phys.~Chem.~Ref.~Data~Suppl.}
\def \jpcrd{J.~Phys.~Chem.~Ref.~Data}
\def \jstqe{IEEE~J.~Sel.~Top.~Quant.~Electron.}
\def \jqe{IEEE~J.~Quant.~Electron.}
\def \jqsrt{J.~Quant.~Spectrosc.~Radiat.~Transfer}
\def \jrnist{J.~Res.~Natl.~Inst.~Stand.~Technol.}
\def \jtp{J.~Technical~Phys.}
\def \lp{Laser~Phys.}
\def \lnp{Lecture~Notes~Phys.}
\def \met{Metrologia}
\def \mnras{Mon.~Not.~R.~Astron.~Soc.}
\def \mp{Mol.~Phys.}
\def \mpla{Mod.~Phys.~Lett.~A}
\def \mst{Meas.~Sci.~Technol.}
\def \nat{Nature}
\def \na{New~Astron.}
\def \nar{New~Astron.~Rev.}
\def \npa{Nucl.~Phys.~A}
\def \npb{Nucl.~Phys.~B}
\def \nsrds{Natl.~Stand.~Rel.~Data~Ser.}
\def \nw{Naturwiss.}
\def \oe{Opt.~Express}
\def \oc{Opt.~Comm.}
\def \ol{Opt.~Lett.}
\def \pasa{Publ.~Astron.~Soc.~Australia}
\def \pasj{Publ.~Astron.~Soc.~Japan}
\def \pasp{Publ.~Astron.~Soc.~Pacific}
\def \pawk{Preuss.~Akad.~Wiss.~K}
\def \planss{Planet. Space Sci.}
\def \plb{Phys.~Lett.~B}
\def \phd{PhD~thesis}
\def \pnas{Proc.~Natl.~Acad.~Sci.}
\def \pra{Phys.~Rev.~A}
\def \prb{Phys.~Rev.~B}
\def \prc{Phys.~Rev.~C}
\def \prd{Phys.~Rev.~D}
\def \prep{in~preparation}
\def \prl{Phys.~Rev.~Lett.}
\def \prsa{Proc.~R.~Soc.~A}
\def \pr{Phys.~Rev.}
\def \psc{Phys.~Scr.}
\def \ptp{Progress~Theor.~Phys.}
\def \ptrsla{Phil.~Trans.~R.~Soc.~London}
\def \rsi{Rev.~Sci.~Instrum.}
\def \rmp{Rev.~Mod.~Phys.}
\def \rpp{Rep.~Prog.~Phys.}
\def \sci{Science}
\def \spjetpl{Sov.~Phys.~JETP~Lett.}
\def \spu{Sov.~Phys.~Uspekhi}
\def \ssr{Space~Sci.~Rev.}
\def \tms{Trans.~Math.~Software}
\def \va{Vistas~Astron.}
\def \za{Z.~Astrophys.}
\def \zpa{Z.~Phys.~A}
\def \zp{Z.~Phys.}

\bibliographystyle{Harvard}
\bibliography{reference}

\begin{thebibliography*}{32}
\providecommand{\bibtype}[1]{}
\providecommand{\natexlab}[1]{#1}
{\catcode`\|=0\catcode`\#=12\catcode`\@=11\catcode`\\=12
|immediate|write|@auxout{\expandafter\ifx\csname natexlab\endcsname\relax\gdef\natexlab#1{#1}\fi}}
\renewcommand{\url}[1]{{\tt #1}}
\providecommand{\urlprefix}{URL }
\expandafter\ifx\csname urlstyle\endcsname\relax
  \providecommand{\doi}[1]{doi:\discretionary{}{}{}#1}\else
  \providecommand{\doi}{doi:\discretionary{}{}{}\begingroup \urlstyle{rm}\Url}\fi
\providecommand{\bibinfo}[2]{#2}
\providecommand{\eprint}[2][]{\url{#2}}

\bibtype{Article}%
\bibitem[{Abbott} et al.(2020)]{DES2020}
\bibinfo{author}{{Abbott} TMC}, \bibinfo{author}{{Aguena} M}, \bibinfo{author}{{Alarcon} A}, \bibinfo{author}{{Allam} S}, \bibinfo{author}{{Allen} S}, \bibinfo{author}{{Annis} J}, \bibinfo{author}{{Avila} S}, \bibinfo{author}{{Bacon} D}, \bibinfo{author}{{Bechtol} K}, \bibinfo{author}{{Bermeo} A}, \bibinfo{author}{{Bernstein} GM}, \bibinfo{author}{{Bertin} E}, \bibinfo{author}{{Bhargava} S}, \bibinfo{author}{{Bocquet} S}, \bibinfo{author}{{Brooks} D}, \bibinfo{author}{{Brout} D}, \bibinfo{author}{{Buckley-Geer} E}, \bibinfo{author}{{Burke} DL}, \bibinfo{author}{{Carnero Rosell} A}, \bibinfo{author}{{Carrasco Kind} M}, \bibinfo{author}{{Carretero} J}, \bibinfo{author}{{Castander} FJ}, \bibinfo{author}{{Cawthon} R}, \bibinfo{author}{{Chang} C}, \bibinfo{author}{{Chen} X}, \bibinfo{author}{{Choi} A}, \bibinfo{author}{{Costanzi} M}, \bibinfo{author}{{Crocce} M}, \bibinfo{author}{{da Costa} LN}, \bibinfo{author}{{Davis} TM}, \bibinfo{author}{{De Vicente} J}, \bibinfo{author}{{DeRose} J}, \bibinfo{author}{{Desai}
  S}, \bibinfo{author}{{Diehl} HT}, \bibinfo{author}{{Dietrich} JP}, \bibinfo{author}{{Dodelson} S}, \bibinfo{author}{{Doel} P}, \bibinfo{author}{{Drlica-Wagner} A}, \bibinfo{author}{{Eckert} K}, \bibinfo{author}{{Eifler} TF}, \bibinfo{author}{{Elvin-Poole} J}, \bibinfo{author}{{Estrada} J}, \bibinfo{author}{{Everett} S}, \bibinfo{author}{{Evrard} AE}, \bibinfo{author}{{Farahi} A}, \bibinfo{author}{{Ferrero} I}, \bibinfo{author}{{Flaugher} B}, \bibinfo{author}{{Fosalba} P}, \bibinfo{author}{{Frieman} J}, \bibinfo{author}{{Garc{\'\i}a-Bellido} J}, \bibinfo{author}{{Gatti} M}, \bibinfo{author}{{Gaztanaga} E}, \bibinfo{author}{{Gerdes} DW}, \bibinfo{author}{{Giannantonio} T}, \bibinfo{author}{{Giles} P}, \bibinfo{author}{{Grandis} S}, \bibinfo{author}{{Gruen} D}, \bibinfo{author}{{Gruendl} RA}, \bibinfo{author}{{Gschwend} J}, \bibinfo{author}{{Gutierrez} G}, \bibinfo{author}{{Hartley} WG}, \bibinfo{author}{{Hinton} SR}, \bibinfo{author}{{Hollowood} DL}, \bibinfo{author}{{Honscheid} K}, \bibinfo{author}{{Hoyle}
  B}, \bibinfo{author}{{Huterer} D}, \bibinfo{author}{{James} DJ}, \bibinfo{author}{{Jarvis} M}, \bibinfo{author}{{Jeltema} T}, \bibinfo{author}{{Johnson} MWG}, \bibinfo{author}{{Johnson} MD}, \bibinfo{author}{{Kent} S}, \bibinfo{author}{{Krause} E}, \bibinfo{author}{{Kron} R}, \bibinfo{author}{{Kuehn} K}, \bibinfo{author}{{Kuropatkin} N}, \bibinfo{author}{{Lahav} O}, \bibinfo{author}{{Li} TS}, \bibinfo{author}{{Lidman} C}, \bibinfo{author}{{Lima} M}, \bibinfo{author}{{Lin} H}, \bibinfo{author}{{MacCrann} N}, \bibinfo{author}{{Maia} MAG}, \bibinfo{author}{{Mantz} A}, \bibinfo{author}{{Marshall} JL}, \bibinfo{author}{{Martini} P}, \bibinfo{author}{{Mayers} J}, \bibinfo{author}{{Melchior} P}, \bibinfo{author}{{Mena-Fern{\'a}ndez} J}, \bibinfo{author}{{Menanteau} F}, \bibinfo{author}{{Miquel} R}, \bibinfo{author}{{Mohr} JJ}, \bibinfo{author}{{Nichol} RC}, \bibinfo{author}{{Nord} B}, \bibinfo{author}{{Ogando} RLC}, \bibinfo{author}{{Palmese} A}, \bibinfo{author}{{Paz-Chinch{\'o}n} F}, \bibinfo{author}{{Plazas}
  AA}, \bibinfo{author}{{Prat} J}, \bibinfo{author}{{Rau} MM}, \bibinfo{author}{{Romer} AK}, \bibinfo{author}{{Roodman} A}, \bibinfo{author}{{Rooney} P}, \bibinfo{author}{{Rozo} E}, \bibinfo{author}{{Rykoff} ES}, \bibinfo{author}{{Sako} M}, \bibinfo{author}{{Samuroff} S}, \bibinfo{author}{{S{\'a}nchez} C}, \bibinfo{author}{{Sanchez} E}, \bibinfo{author}{{Saro} A}, \bibinfo{author}{{Scarpine} V}, \bibinfo{author}{{Schubnell} M}, \bibinfo{author}{{Scolnic} D}, \bibinfo{author}{{Serrano} S}, \bibinfo{author}{{Sevilla-Noarbe} I}, \bibinfo{author}{{Sheldon} E}, \bibinfo{author}{{Smith} JA}, \bibinfo{author}{{Smith} M}, \bibinfo{author}{{Suchyta} E}, \bibinfo{author}{{Swanson} MEC}, \bibinfo{author}{{Tarle} G}, \bibinfo{author}{{Thomas} D}, \bibinfo{author}{{To} C}, \bibinfo{author}{{Troxel} MA}, \bibinfo{author}{{Tucker} DL}, \bibinfo{author}{{Varga} TN}, \bibinfo{author}{{von der Linden} A}, \bibinfo{author}{{Walker} AR}, \bibinfo{author}{{Wechsler} RH}, \bibinfo{author}{{Weller} J}, \bibinfo{author}{{Wilkinson}
  RD}, \bibinfo{author}{{Wu} H}, \bibinfo{author}{{Yanny} B}, \bibinfo{author}{{Zhang} Y}, \bibinfo{author}{{Zhang} Z}, \bibinfo{author}{{Zuntz} J} and  \bibinfo{author}{{DES Collaboration}} (\bibinfo{year}{2020}), \bibinfo{month}{Jul.}
\bibinfo{title}{{Dark Energy Survey Year 1 Results: Cosmological constraints from cluster abundances and weak lensing}}.
\bibinfo{journal}{{\em \prd}} \bibinfo{volume}{102} (\bibinfo{number}{2}), \bibinfo{eid}{023509}. \bibinfo{doi}{\doi{10.1103/PhysRevD.102.023509}}.
\eprint{2002.11124}.

\bibtype{Article}%
\bibitem[{Alam} et al.(2021)]{BOSS2021}
\bibinfo{author}{{Alam} S}, \bibinfo{author}{{Aubert} M}, \bibinfo{author}{{Avila} S}, \bibinfo{author}{{Balland} C}, \bibinfo{author}{{Bautista} JE}, \bibinfo{author}{{Bershady} MA}, \bibinfo{author}{{Bizyaev} D}, \bibinfo{author}{{Blanton} MR}, \bibinfo{author}{{Bolton} AS}, \bibinfo{author}{{Bovy} J}, \bibinfo{author}{{Brinkmann} J}, \bibinfo{author}{{Brownstein} JR}, \bibinfo{author}{{Burtin} E}, \bibinfo{author}{{Chabanier} S}, \bibinfo{author}{{Chapman} MJ}, \bibinfo{author}{{Choi} PD}, \bibinfo{author}{{Chuang} CH}, \bibinfo{author}{{Comparat} J}, \bibinfo{author}{{Cousinou} MC}, \bibinfo{author}{{Cuceu} A}, \bibinfo{author}{{Dawson} KS}, \bibinfo{author}{{de la Torre} S}, \bibinfo{author}{{de Mattia} A}, \bibinfo{author}{{Agathe} VdS}, \bibinfo{author}{{des Bourboux} HdM}, \bibinfo{author}{{Escoffier} S}, \bibinfo{author}{{Etourneau} T}, \bibinfo{author}{{Farr} J}, \bibinfo{author}{{Font-Ribera} A}, \bibinfo{author}{{Frinchaboy} PM}, \bibinfo{author}{{Fromenteau} S}, \bibinfo{author}{{Gil-Mar{\'\i}n}
  H}, \bibinfo{author}{{Le Goff} JM}, \bibinfo{author}{{Gonzalez-Morales} AX}, \bibinfo{author}{{Gonzalez-Perez} V}, \bibinfo{author}{{Grabowski} K}, \bibinfo{author}{{Guy} J}, \bibinfo{author}{{Hawken} AJ}, \bibinfo{author}{{Hou} J}, \bibinfo{author}{{Kong} H}, \bibinfo{author}{{Parker} J}, \bibinfo{author}{{Klaene} M}, \bibinfo{author}{{Kneib} JP}, \bibinfo{author}{{Lin} S}, \bibinfo{author}{{Long} D}, \bibinfo{author}{{Lyke} BW}, \bibinfo{author}{{de la Macorra} A}, \bibinfo{author}{{Martini} P}, \bibinfo{author}{{Masters} K}, \bibinfo{author}{{Mohammad} FG}, \bibinfo{author}{{Moon} J}, \bibinfo{author}{{Mueller} EM}, \bibinfo{author}{{Mu{\~n}oz-Guti{\'e}rrez} A}, \bibinfo{author}{{Myers} AD}, \bibinfo{author}{{Nadathur} S}, \bibinfo{author}{{Neveux} R}, \bibinfo{author}{{Newman} JA}, \bibinfo{author}{{Noterdaeme} P}, \bibinfo{author}{{Oravetz} A}, \bibinfo{author}{{Oravetz} D}, \bibinfo{author}{{Palanque-Delabrouille} N}, \bibinfo{author}{{Pan} K}, \bibinfo{author}{{Paviot} R}, \bibinfo{author}{{Percival}
  WJ}, \bibinfo{author}{{P{\'e}rez-R{\`a}fols} I}, \bibinfo{author}{{Petitjean} P}, \bibinfo{author}{{Pieri} MM}, \bibinfo{author}{{Prakash} A}, \bibinfo{author}{{Raichoor} A}, \bibinfo{author}{{Ravoux} C}, \bibinfo{author}{{Rezaie} M}, \bibinfo{author}{{Rich} J}, \bibinfo{author}{{Ross} AJ}, \bibinfo{author}{{Rossi} G}, \bibinfo{author}{{Ruggeri} R}, \bibinfo{author}{{Ruhlmann-Kleider} V}, \bibinfo{author}{{S{\'a}nchez} AG}, \bibinfo{author}{{S{\'a}nchez} FJ}, \bibinfo{author}{{S{\'a}nchez-Gallego} JR}, \bibinfo{author}{{Sayres} C}, \bibinfo{author}{{Schneider} DP}, \bibinfo{author}{{Seo} HJ}, \bibinfo{author}{{Shafieloo} A}, \bibinfo{author}{{Slosar} A}, \bibinfo{author}{{Smith} A}, \bibinfo{author}{{Stermer} J}, \bibinfo{author}{{Tamone} A}, \bibinfo{author}{{Tinker} JL}, \bibinfo{author}{{Tojeiro} R}, \bibinfo{author}{{Vargas-Maga{\~n}a} M}, \bibinfo{author}{{Variu} A}, \bibinfo{author}{{Wang} Y}, \bibinfo{author}{{Weaver} BA}, \bibinfo{author}{{Weijmans} AM}, \bibinfo{author}{{Y{\`e}che} C},
  \bibinfo{author}{{Zarrouk} P}, \bibinfo{author}{{Zhao} C}, \bibinfo{author}{{Zhao} GB} and  \bibinfo{author}{{Zheng} Z} (\bibinfo{year}{2021}), \bibinfo{month}{Apr.}
\bibinfo{title}{{Completed SDSS-IV extended Baryon Oscillation Spectroscopic Survey: Cosmological implications from two decades of spectroscopic surveys at the Apache Point Observatory}}.
\bibinfo{journal}{{\em \prd}} \bibinfo{volume}{103} (\bibinfo{number}{8}), \bibinfo{eid}{083533}. \bibinfo{doi}{\doi{10.1103/PhysRevD.103.083533}}.
\eprint{2007.08991}.

\bibtype{Article}%
\bibitem[{Asgari} et al.(2021)]{KIDS2021}
\bibinfo{author}{{Asgari} M}, \bibinfo{author}{{Lin} CA}, \bibinfo{author}{{Joachimi} B}, \bibinfo{author}{{Giblin} B}, \bibinfo{author}{{Heymans} C}, \bibinfo{author}{{Hildebrandt} H}, \bibinfo{author}{{Kannawadi} A}, \bibinfo{author}{{St{\"o}lzner} B}, \bibinfo{author}{{Tr{\"o}ster} T}, \bibinfo{author}{{van den Busch} JL}, \bibinfo{author}{{Wright} AH}, \bibinfo{author}{{Bilicki} M}, \bibinfo{author}{{Blake} C}, \bibinfo{author}{{de Jong} J}, \bibinfo{author}{{Dvornik} A}, \bibinfo{author}{{Erben} T}, \bibinfo{author}{{Getman} F}, \bibinfo{author}{{Hoekstra} H}, \bibinfo{author}{{K{\"o}hlinger} F}, \bibinfo{author}{{Kuijken} K}, \bibinfo{author}{{Miller} L}, \bibinfo{author}{{Radovich} M}, \bibinfo{author}{{Schneider} P}, \bibinfo{author}{{Shan} H} and  \bibinfo{author}{{Valentijn} E} (\bibinfo{year}{2021}), \bibinfo{month}{Jan.}
\bibinfo{title}{{KiDS-1000 cosmology: Cosmic shear constraints and comparison between two point statistics}}.
\bibinfo{journal}{{\em \aap}} \bibinfo{volume}{645}, \bibinfo{eid}{A104}. \bibinfo{doi}{\doi{10.1051/0004-6361/202039070}}.
\eprint{2007.15633}.

\bibtype{Article}%
\bibitem[{Baumann} et al.(2012)]{Baumann2012}
\bibinfo{author}{{Baumann} D}, \bibinfo{author}{{Nicolis} A}, \bibinfo{author}{{Senatore} L} and  \bibinfo{author}{{Zaldarriaga} M} (\bibinfo{year}{2012}), \bibinfo{month}{Jul.}
\bibinfo{title}{{Cosmological non-linearities as an effective fluid}}.
\bibinfo{journal}{{\em \jcap}} \bibinfo{volume}{2012} (\bibinfo{number}{7}), \bibinfo{eid}{051}. \bibinfo{doi}{\doi{10.1088/1475-7516/2012/07/051}}.
\eprint{1004.2488}.

\bibtype{Article}%
\bibitem[{Buchert}(2008)]{Buchert2008}
\bibinfo{author}{{Buchert} T} (\bibinfo{year}{2008}), \bibinfo{month}{Feb.}
\bibinfo{title}{{Dark Energy from structure: a status report}}.
\bibinfo{journal}{{\em General Relativity and Gravitation}} \bibinfo{volume}{40} (\bibinfo{number}{2-3}): \bibinfo{pages}{467--527}. \bibinfo{doi}{\doi{10.1007/s10714-007-0554-8}}.
\eprint{0707.2153}.

\bibtype{Book}%
\bibitem[{Carroll}(2004)]{Carroll2004}
\bibinfo{author}{{Carroll} SM} (\bibinfo{year}{2004}).
\bibinfo{title}{{Spacetime and Geometry, An Introduction to General Relativity}}, \bibinfo{address}{Pearson Education, San Francisco}.
\bibinfo{pages}{Chapter 8} .

\bibtype{Article}%
\bibitem[{Carroll} et al.(1992)]{Carroll1992}
\bibinfo{author}{{Carroll} SM}, \bibinfo{author}{{Press} WH} and  \bibinfo{author}{{Turner} EL} (\bibinfo{year}{1992}), \bibinfo{month}{Jan.}
\bibinfo{title}{{The cosmological constant.}}
\bibinfo{journal}{{\em \araa}} \bibinfo{volume}{30}: \bibinfo{pages}{499--542}. \bibinfo{doi}{\doi{10.1146/annurev.aa.30.090192.002435}}.

\bibtype{Article}%
\bibitem[{Chevallier} and {Polarski}(2001)]{Chevallier2001}
\bibinfo{author}{{Chevallier} M} and  \bibinfo{author}{{Polarski} D} (\bibinfo{year}{2001}), \bibinfo{month}{Jan.}
\bibinfo{title}{{Accelerating Universes with Scaling Dark Matter}}.
\bibinfo{journal}{{\em International Journal of Modern Physics D}} \bibinfo{volume}{10} (\bibinfo{number}{2}): \bibinfo{pages}{213--223}. \bibinfo{doi}{\doi{10.1142/S0218271801000822}}.
\eprint{gr-qc/0009008}.

\bibtype{Article}%
\bibitem[{Davis} et al.(2019)]{Davis2019}
\bibinfo{author}{{Davis} TM}, \bibinfo{author}{{Hinton} SR}, \bibinfo{author}{{Howlett} C} and  \bibinfo{author}{{Calcino} J} (\bibinfo{year}{2019}), \bibinfo{month}{Dec.}
\bibinfo{title}{{Can redshift errors bias measurements of the Hubble Constant?}}
\bibinfo{journal}{{\em \mnras}} \bibinfo{volume}{490} (\bibinfo{number}{2}): \bibinfo{pages}{2948--2957}. \bibinfo{doi}{\doi{10.1093/mnras/stz2652}}.
\eprint{1907.12639}.

\bibtype{Article}%
\bibitem[{DES Collaboration} et al.(2024)]{des2024}
\bibinfo{author}{{DES Collaboration}}, \bibinfo{author}{{Abbott} TMC}, \bibinfo{author}{{Acevedo} M}, \bibinfo{author}{{Aguena} M}, \bibinfo{author}{{Alarcon} A}, \bibinfo{author}{{Allam} S}, \bibinfo{author}{{Alves} O}, \bibinfo{author}{{Amon} A}, \bibinfo{author}{{Andrade-Oliveira} F}, \bibinfo{author}{{Annis} J}, \bibinfo{author}{{Armstrong} P}, \bibinfo{author}{{Asorey} J}, \bibinfo{author}{{Avila} S}, \bibinfo{author}{{Bacon} D}, \bibinfo{author}{{Bassett} BA}, \bibinfo{author}{{Bechtol} K}, \bibinfo{author}{{Bernardinelli} PH}, \bibinfo{author}{{Bernstein} GM}, \bibinfo{author}{{Bertin} E}, \bibinfo{author}{{Blazek} J}, \bibinfo{author}{{Bocquet} S}, \bibinfo{author}{{Brooks} D}, \bibinfo{author}{{Brout} D}, \bibinfo{author}{{Buckley-Geer} E}, \bibinfo{author}{{Burke} DL}, \bibinfo{author}{{Camacho} H}, \bibinfo{author}{{Camilleri} R}, \bibinfo{author}{{Campos} A}, \bibinfo{author}{{Carnero Rosell} A}, \bibinfo{author}{{Carollo} D}, \bibinfo{author}{{Carr} A}, \bibinfo{author}{{Carretero} J},
  \bibinfo{author}{{Castander} FJ}, \bibinfo{author}{{Cawthon} R}, \bibinfo{author}{{Chang} C}, \bibinfo{author}{{Chen} R}, \bibinfo{author}{{Choi} A}, \bibinfo{author}{{Conselice} C}, \bibinfo{author}{{Costanzi} M}, \bibinfo{author}{{da Costa} LN}, \bibinfo{author}{{Crocce} M}, \bibinfo{author}{{Davis} TM}, \bibinfo{author}{{DePoy} DL}, \bibinfo{author}{{Desai} S}, \bibinfo{author}{{Diehl} HT}, \bibinfo{author}{{Dixon} M}, \bibinfo{author}{{Dodelson} S}, \bibinfo{author}{{Doel} P}, \bibinfo{author}{{Doux} C}, \bibinfo{author}{{Drlica-Wagner} A}, \bibinfo{author}{{Elvin-Poole} J}, \bibinfo{author}{{Everett} S}, \bibinfo{author}{{Ferrero} I}, \bibinfo{author}{{Fert{\'e}} A}, \bibinfo{author}{{Flaugher} B}, \bibinfo{author}{{Foley} RJ}, \bibinfo{author}{{Fosalba} P}, \bibinfo{author}{{Friedel} D}, \bibinfo{author}{{Frieman} J}, \bibinfo{author}{{Frohmaier} C}, \bibinfo{author}{{Galbany} L}, \bibinfo{author}{{Garc{\'\i}a-Bellido} J}, \bibinfo{author}{{Gatti} M}, \bibinfo{author}{{Gaztanaga} E},
  \bibinfo{author}{{Giannini} G}, \bibinfo{author}{{Glazebrook} K}, \bibinfo{author}{{Graur} O}, \bibinfo{author}{{Gruen} D}, \bibinfo{author}{{Gruendl} RA}, \bibinfo{author}{{Gutierrez} G}, \bibinfo{author}{{Hartley} WG}, \bibinfo{author}{{Herner} K}, \bibinfo{author}{{Hinton} SR}, \bibinfo{author}{{Hollowood} DL}, \bibinfo{author}{{Honscheid} K}, \bibinfo{author}{{Huterer} D}, \bibinfo{author}{{Jain} B}, \bibinfo{author}{{James} DJ}, \bibinfo{author}{{Jeffrey} N}, \bibinfo{author}{{Kasai} E}, \bibinfo{author}{{Kelsey} L}, \bibinfo{author}{{Kent} S}, \bibinfo{author}{{Kessler} R}, \bibinfo{author}{{Kim} AG}, \bibinfo{author}{{Kirshner} RP}, \bibinfo{author}{{Kovacs} E}, \bibinfo{author}{{Kuehn} K}, \bibinfo{author}{{Lahav} O}, \bibinfo{author}{{Lee} J}, \bibinfo{author}{{Lee} S}, \bibinfo{author}{{Lewis} GF}, \bibinfo{author}{{Li} TS}, \bibinfo{author}{{Lidman} C}, \bibinfo{author}{{Lin} H}, \bibinfo{author}{{Malik} U}, \bibinfo{author}{{Marshall} JL}, \bibinfo{author}{{Martini} P},
  \bibinfo{author}{{Mena-Fern{\'a}ndez} J}, \bibinfo{author}{{Menanteau} F}, \bibinfo{author}{{Miquel} R}, \bibinfo{author}{{Mohr} JJ}, \bibinfo{author}{{Mould} J}, \bibinfo{author}{{Muir} J}, \bibinfo{author}{{M{\"o}ller} A}, \bibinfo{author}{{Neilsen} E}, \bibinfo{author}{{Nichol} RC}, \bibinfo{author}{{Nugent} P}, \bibinfo{author}{{Ogando} RLC}, \bibinfo{author}{{Palmese} A}, \bibinfo{author}{{Pan} YC}, \bibinfo{author}{{Paterno} M}, \bibinfo{author}{{Percival} WJ}, \bibinfo{author}{{Pereira} MES}, \bibinfo{author}{{Pieres} A}, \bibinfo{author}{{Malag{\'o}n} AAP}, \bibinfo{author}{{Popovic} B}, \bibinfo{author}{{Porredon} A}, \bibinfo{author}{{Prat} J}, \bibinfo{author}{{Qu} H}, \bibinfo{author}{{Raveri} M}, \bibinfo{author}{{Rodr{\'\i}guez-Monroy} M}, \bibinfo{author}{{Romer} AK}, \bibinfo{author}{{Roodman} A}, \bibinfo{author}{{Rose} B}, \bibinfo{author}{{Sako} M}, \bibinfo{author}{{Sanchez} E}, \bibinfo{author}{{Sanchez Cid} D}, \bibinfo{author}{{Schubnell} M}, \bibinfo{author}{{Scolnic} D},
  \bibinfo{author}{{Sevilla-Noarbe} I}, \bibinfo{author}{{Shah} P}, \bibinfo{author}{{Smith} JA}, \bibinfo{author}{{Smith} M}, \bibinfo{author}{{Soares-Santos} M}, \bibinfo{author}{{Suchyta} E}, \bibinfo{author}{{Sullivan} M}, \bibinfo{author}{{Suntzeff} N}, \bibinfo{author}{{Swanson} MEC}, \bibinfo{author}{{S{\'a}nchez} BO}, \bibinfo{author}{{Tarle} G}, \bibinfo{author}{{Taylor} G}, \bibinfo{author}{{Thomas} D}, \bibinfo{author}{{To} C}, \bibinfo{author}{{Toy} M}, \bibinfo{author}{{Troxel} MA}, \bibinfo{author}{{Tucker} BE}, \bibinfo{author}{{Tucker} DL}, \bibinfo{author}{{Uddin} SA}, \bibinfo{author}{{Vincenzi} M}, \bibinfo{author}{{Walker} AR}, \bibinfo{author}{{Weaverdyck} N}, \bibinfo{author}{{Wechsler} RH}, \bibinfo{author}{{Weller} J}, \bibinfo{author}{{Wester} W}, \bibinfo{author}{{Wiseman} P}, \bibinfo{author}{{Yamamoto} M}, \bibinfo{author}{{Yuan} F}, \bibinfo{author}{{Zhang} B} and  \bibinfo{author}{{Zhang} Y} (\bibinfo{year}{2024}), \bibinfo{month}{Sep.}
\bibinfo{title}{{The Dark Energy Survey: Cosmology Results with {\ensuremath{\sim}}1500 New High-redshift Type Ia Supernovae Using the Full 5 yr Data Set}}.
\bibinfo{journal}{{\em \apjl}} \bibinfo{volume}{973} (\bibinfo{number}{1}), \bibinfo{eid}{L14}. \bibinfo{doi}{\doi{10.3847/2041-8213/ad6f9f}}.
\eprint{2401.02929}.

\bibtype{Article}%
\bibitem[{DESI Collaboration} et al.(2024)]{desi2024}
\bibinfo{author}{{DESI Collaboration}}, \bibinfo{author}{{Adame} AG}, \bibinfo{author}{{Aguilar} J}, \bibinfo{author}{{Ahlen} S}, \bibinfo{author}{{Alam} S}, \bibinfo{author}{{Alexander} DM}, \bibinfo{author}{{Alvarez} M}, \bibinfo{author}{{Alves} O}, \bibinfo{author}{{Anand} A}, \bibinfo{author}{{Andrade} U}, \bibinfo{author}{{Armengaud} E}, \bibinfo{author}{{Avila} S}, \bibinfo{author}{{Aviles} A}, \bibinfo{author}{{Awan} H}, \bibinfo{author}{{Bahr-Kalus} B}, \bibinfo{author}{{Bailey} S}, \bibinfo{author}{{Baltay} C}, \bibinfo{author}{{Bault} A}, \bibinfo{author}{{Behera} J}, \bibinfo{author}{{BenZvi} S}, \bibinfo{author}{{Bera} A}, \bibinfo{author}{{Beutler} F}, \bibinfo{author}{{Bianchi} D}, \bibinfo{author}{{Blake} C}, \bibinfo{author}{{Blum} R}, \bibinfo{author}{{Brieden} S}, \bibinfo{author}{{Brodzeller} A}, \bibinfo{author}{{Brooks} D}, \bibinfo{author}{{Buckley-Geer} E}, \bibinfo{author}{{Burtin} E}, \bibinfo{author}{{Calderon} R}, \bibinfo{author}{{Canning} R}, \bibinfo{author}{{Carnero Rosell} A},
  \bibinfo{author}{{Cereskaite} R}, \bibinfo{author}{{Cervantes-Cota} JL}, \bibinfo{author}{{Chabanier} S}, \bibinfo{author}{{Chaussidon} E}, \bibinfo{author}{{Chaves-Montero} J}, \bibinfo{author}{{Chen} S}, \bibinfo{author}{{Chen} X}, \bibinfo{author}{{Claybaugh} T}, \bibinfo{author}{{Cole} S}, \bibinfo{author}{{Cuceu} A}, \bibinfo{author}{{Davis} TM}, \bibinfo{author}{{Dawson} K}, \bibinfo{author}{{de la Macorra} A}, \bibinfo{author}{{de Mattia} A}, \bibinfo{author}{{Deiosso} N}, \bibinfo{author}{{Dey} A}, \bibinfo{author}{{Dey} B}, \bibinfo{author}{{Ding} Z}, \bibinfo{author}{{Doel} P}, \bibinfo{author}{{Edelstein} J}, \bibinfo{author}{{Eftekharzadeh} S}, \bibinfo{author}{{Eisenstein} DJ}, \bibinfo{author}{{Elliott} A}, \bibinfo{author}{{Fagrelius} P}, \bibinfo{author}{{Fanning} K}, \bibinfo{author}{{Ferraro} S}, \bibinfo{author}{{Ereza} J}, \bibinfo{author}{{Findlay} N}, \bibinfo{author}{{Flaugher} B}, \bibinfo{author}{{Font-Ribera} A}, \bibinfo{author}{{Forero-S{\'a}nchez} D},
  \bibinfo{author}{{Forero-Romero} JE}, \bibinfo{author}{{Frenk} CS}, \bibinfo{author}{{Garcia-Quintero} C}, \bibinfo{author}{{Gazta{\~n}aga} E}, \bibinfo{author}{{Gil-Mar{\'\i}n} H}, \bibinfo{author}{{Gontcho} SGA}, \bibinfo{author}{{Gonzalez-Morales} AX}, \bibinfo{author}{{Gonzalez-Perez} V}, \bibinfo{author}{{Gordon} C}, \bibinfo{author}{{Green} D}, \bibinfo{author}{{Gruen} D}, \bibinfo{author}{{Gsponer} R}, \bibinfo{author}{{Gutierrez} G}, \bibinfo{author}{{Guy} J}, \bibinfo{author}{{Hadzhiyska} B}, \bibinfo{author}{{Hahn} C}, \bibinfo{author}{{Hanif} MMS}, \bibinfo{author}{{Herrera-Alcantar} HK}, \bibinfo{author}{{Honscheid} K}, \bibinfo{author}{{Howlett} C}, \bibinfo{author}{{Huterer} D}, \bibinfo{author}{{Ir{\v{s}}i{\v{c}}} V}, \bibinfo{author}{{Ishak} M}, \bibinfo{author}{{Juneau} S}, \bibinfo{author}{{Kara{\c{c}}ayl{\i}} NG}, \bibinfo{author}{{Kehoe} R}, \bibinfo{author}{{Kent} S}, \bibinfo{author}{{Kirkby} D}, \bibinfo{author}{{Kremin} A}, \bibinfo{author}{{Krolewski} A}, \bibinfo{author}{{Lai} Y},
  \bibinfo{author}{{Lan} TW}, \bibinfo{author}{{Landriau} M}, \bibinfo{author}{{Lang} D}, \bibinfo{author}{{Lasker} J}, \bibinfo{author}{{Le Goff} JM}, \bibinfo{author}{{Le Guillou} L}, \bibinfo{author}{{Leauthaud} A}, \bibinfo{author}{{Levi} ME}, \bibinfo{author}{{Li} TS}, \bibinfo{author}{{Linder} E}, \bibinfo{author}{{Lodha} K}, \bibinfo{author}{{Magneville} C}, \bibinfo{author}{{Manera} M}, \bibinfo{author}{{Margala} D}, \bibinfo{author}{{Martini} P}, \bibinfo{author}{{Maus} M}, \bibinfo{author}{{McDonald} P}, \bibinfo{author}{{Medina-Varela} L}, \bibinfo{author}{{Meisner} A}, \bibinfo{author}{{Mena-Fern{\'a}ndez} J}, \bibinfo{author}{{Miquel} R}, \bibinfo{author}{{Moon} J}, \bibinfo{author}{{Moore} S}, \bibinfo{author}{{Moustakas} J}, \bibinfo{author}{{Mudur} N}, \bibinfo{author}{{Mueller} E}, \bibinfo{author}{{Mu{\~n}oz-Guti{\'e}rrez} A}, \bibinfo{author}{{Myers} AD}, \bibinfo{author}{{Nadathur} S}, \bibinfo{author}{{Napolitano} L}, \bibinfo{author}{{Neveux} R}, \bibinfo{author}{{Newman} JA},
  \bibinfo{author}{{Nguyen} NM}, \bibinfo{author}{{Nie} J}, \bibinfo{author}{{Niz} G}, \bibinfo{author}{{Noriega} HE}, \bibinfo{author}{{Padmanabhan} N}, \bibinfo{author}{{Paillas} E}, \bibinfo{author}{{Palanque-Delabrouille} N}, \bibinfo{author}{{Pan} J}, \bibinfo{author}{{Penmetsa} S}, \bibinfo{author}{{Percival} WJ}, \bibinfo{author}{{Pieri} MM}, \bibinfo{author}{{Pinon} M}, \bibinfo{author}{{Poppett} C}, \bibinfo{author}{{Porredon} A}, \bibinfo{author}{{Prada} F}, \bibinfo{author}{{P{\'e}rez-Fern{\'a}ndez} A}, \bibinfo{author}{{P{\'e}rez-R{\`a}fols} I}, \bibinfo{author}{{Rabinowitz} D}, \bibinfo{author}{{Raichoor} A}, \bibinfo{author}{{Ram{\'\i}rez-P{\'e}rez} C}, \bibinfo{author}{{Ramirez-Solano} S}, \bibinfo{author}{{Ravoux} C}, \bibinfo{author}{{Rashkovetskyi} M}, \bibinfo{author}{{Rezaie} M}, \bibinfo{author}{{Rich} J}, \bibinfo{author}{{Rocher} A}, \bibinfo{author}{{Rockosi} C}, \bibinfo{author}{{Roe} NA}, \bibinfo{author}{{Rosado-Marin} A}, \bibinfo{author}{{Ross} AJ}, \bibinfo{author}{{Rossi} G},
  \bibinfo{author}{{Ruggeri} R}, \bibinfo{author}{{Ruhlmann-Kleider} V}, \bibinfo{author}{{Samushia} L}, \bibinfo{author}{{Sanchez} E}, \bibinfo{author}{{Saulder} C}, \bibinfo{author}{{Schlafly} EF}, \bibinfo{author}{{Schlegel} D}, \bibinfo{author}{{Schubnell} M}, \bibinfo{author}{{Seo} H}, \bibinfo{author}{{Shafieloo} A}, \bibinfo{author}{{Sharples} R}, \bibinfo{author}{{Silber} J}, \bibinfo{author}{{Slosar} A}, \bibinfo{author}{{Smith} A}, \bibinfo{author}{{Sprayberry} D}, \bibinfo{author}{{Tan} T}, \bibinfo{author}{{Tarl{\'e}} G}, \bibinfo{author}{{Taylor} P}, \bibinfo{author}{{Trusov} S}, \bibinfo{author}{{Ure{\~n}a-L{\'o}pez} LA}, \bibinfo{author}{{Vaisakh} R}, \bibinfo{author}{{Valcin} D}, \bibinfo{author}{{Valdes} F}, \bibinfo{author}{{Vargas-Maga{\~n}a} M}, \bibinfo{author}{{Verde} L}, \bibinfo{author}{{Walther} M}, \bibinfo{author}{{Wang} B}, \bibinfo{author}{{Wang} MS}, \bibinfo{author}{{Weaver} BA}, \bibinfo{author}{{Weaverdyck} N}, \bibinfo{author}{{Wechsler} RH}, \bibinfo{author}{{Weinberg} DH},
  \bibinfo{author}{{White} M}, \bibinfo{author}{{Yu} J}, \bibinfo{author}{{Yu} Y}, \bibinfo{author}{{Yuan} S}, \bibinfo{author}{{Y{\`e}che} C}, \bibinfo{author}{{Zaborowski} EA}, \bibinfo{author}{{Zarrouk} P}, \bibinfo{author}{{Zhang} H}, \bibinfo{author}{{Zhao} C} and  \bibinfo{author}{{Zhao} R} (\bibinfo{year}{2024}), \bibinfo{month}{Apr.}
\bibinfo{title}{{DESI 2024 VI: Cosmological Constraints from the Measurements of Baryon Acoustic Oscillations}}.
\bibinfo{journal}{{\em arXiv e-prints}} , \bibinfo{eid}{arXiv:2404.03002}\bibinfo{doi}{\doi{10.48550/arXiv.2404.03002}}.
\eprint{2404.03002}.

\bibtype{Article}%
\bibitem[{Di Valentino} et al.(2021)]{Valentino2021}
\bibinfo{author}{{Di Valentino} E}, \bibinfo{author}{{Mena} O}, \bibinfo{author}{{Pan} S}, \bibinfo{author}{{Visinelli} L}, \bibinfo{author}{{Yang} W}, \bibinfo{author}{{Melchiorri} A}, \bibinfo{author}{{Mota} DF}, \bibinfo{author}{{Riess} AG} and  \bibinfo{author}{{Silk} J} (\bibinfo{year}{2021}), \bibinfo{month}{Jul.}
\bibinfo{title}{{In the realm of the Hubble tension-a review of solutions}}.
\bibinfo{journal}{{\em Classical and Quantum Gravity}} \bibinfo{volume}{38} (\bibinfo{number}{15}), \bibinfo{eid}{153001}. \bibinfo{doi}{\doi{10.1088/1361-6382/ac086d}}.
\eprint{2103.01183}.

\bibtype{Article}%
\bibitem[{Etherington}(1933)]{Etherington1933}
\bibinfo{author}{{Etherington} IMH} (\bibinfo{year}{1933}), \bibinfo{month}{Jan.}
\bibinfo{title}{{On the Definition of Distance in General Relativity.}}
\bibinfo{journal}{{\em Philosophical Magazine}} \bibinfo{volume}{15} (\bibinfo{number}{18}): \bibinfo{pages}{761}.

\bibtype{Article}%
\bibitem[{Giani} et al.(2024)]{Giani2024}
\bibinfo{author}{{Giani} L}, \bibinfo{author}{{Von Marttens} R} and  \bibinfo{author}{{Camilleri} R} (\bibinfo{year}{2024}), \bibinfo{month}{Oct.}
\bibinfo{title}{{A novel approach to cosmological non-linearities as an effective fluid}}.
\bibinfo{journal}{{\em arXiv e-prints}} , \bibinfo{eid}{arXiv:2410.15295}\bibinfo{doi}{\doi{10.48550/arXiv.2410.15295}}.
\eprint{2410.15295}.

\bibtype{Book}%
\bibitem[{Hartle}(2003)]{Hartle2003}
\bibinfo{author}{{Hartle} JB} (\bibinfo{year}{2003}).
\bibinfo{title}{{Gravity : an introduction to Einstein's general relativity}}, \bibinfo{address}{Addison Wesley, San Francisco}.

\bibtype{Article}%
\bibitem[{Hikage} et al.(2019)]{Hikage2019}
\bibinfo{author}{{Hikage} C}, \bibinfo{author}{{Oguri} M}, \bibinfo{author}{{Hamana} T}, \bibinfo{author}{{More} S}, \bibinfo{author}{{Mandelbaum} R}, \bibinfo{author}{{Takada} M}, \bibinfo{author}{{K{\"o}hlinger} F}, \bibinfo{author}{{Miyatake} H}, \bibinfo{author}{{Nishizawa} AJ}, \bibinfo{author}{{Aihara} H}, \bibinfo{author}{{Armstrong} R}, \bibinfo{author}{{Bosch} J}, \bibinfo{author}{{Coupon} J}, \bibinfo{author}{{Ducout} A}, \bibinfo{author}{{Ho} P}, \bibinfo{author}{{Hsieh} BC}, \bibinfo{author}{{Komiyama} Y}, \bibinfo{author}{{Lanusse} F}, \bibinfo{author}{{Leauthaud} A}, \bibinfo{author}{{Lupton} RH}, \bibinfo{author}{{Medezinski} E}, \bibinfo{author}{{Mineo} S}, \bibinfo{author}{{Miyama} S}, \bibinfo{author}{{Miyazaki} S}, \bibinfo{author}{{Murata} R}, \bibinfo{author}{{Murayama} H}, \bibinfo{author}{{Shirasaki} M}, \bibinfo{author}{{Sif{\'o}n} C}, \bibinfo{author}{{Simet} M}, \bibinfo{author}{{Speagle} J}, \bibinfo{author}{{Spergel} DN}, \bibinfo{author}{{Strauss} MA}, \bibinfo{author}{{Sugiyama}
  N}, \bibinfo{author}{{Tanaka} M}, \bibinfo{author}{{Utsumi} Y}, \bibinfo{author}{{Wang} SY} and  \bibinfo{author}{{Yamada} Y} (\bibinfo{year}{2019}), \bibinfo{month}{Apr.}
\bibinfo{title}{{Cosmology from cosmic shear power spectra with Subaru Hyper Suprime-Cam first-year data}}.
\bibinfo{journal}{{\em \pasj}} \bibinfo{volume}{71} (\bibinfo{number}{2}), \bibinfo{eid}{43}. \bibinfo{doi}{\doi{10.1093/pasj/psz010}}.
\eprint{1809.09148}.

\bibtype{Article}%
\bibitem[{Hubble}(1929)]{Hubble1929}
\bibinfo{author}{{Hubble} E} (\bibinfo{year}{1929}), \bibinfo{month}{Mar.}
\bibinfo{title}{{A Relation between Distance and Radial Velocity among Extra-Galactic Nebulae}}.
\bibinfo{journal}{{\em Proceedings of the National Academy of Science}} \bibinfo{volume}{15} (\bibinfo{number}{3}): \bibinfo{pages}{168--173}. \bibinfo{doi}{\doi{10.1073/pnas.15.3.168}}.

\bibtype{Article}%
\bibitem[{Ishibashi} and {Wald}(2006)]{Ishibashi2006}
\bibinfo{author}{{Ishibashi} A} and  \bibinfo{author}{{Wald} RM} (\bibinfo{year}{2006}), \bibinfo{month}{Jan.}
\bibinfo{title}{{Can the acceleration of our universe be explained by the effects of inhomogeneities?}}
\bibinfo{journal}{{\em Classical and Quantum Gravity}} \bibinfo{volume}{23} (\bibinfo{number}{1}): \bibinfo{pages}{235--250}. \bibinfo{doi}{\doi{10.1088/0264-9381/23/1/012}}.
\eprint{gr-qc/0509108}.

\bibtype{Article}%
\bibitem[{Kaiser}(2017)]{Kaiser2017}
\bibinfo{author}{{Kaiser} N} (\bibinfo{year}{2017}), \bibinfo{month}{Jul.}
\bibinfo{title}{{Why there is no Newtonian backreaction}}.
\bibinfo{journal}{{\em \mnras}} \bibinfo{volume}{469} (\bibinfo{number}{1}): \bibinfo{pages}{744--748}. \bibinfo{doi}{\doi{10.1093/mnras/stx907}}.
\eprint{1703.08809}.

\bibtype{Article}%
\bibitem[{Lema{\^\i}tre}(1927)]{Lemaitre1927}
\bibinfo{author}{{Lema{\^\i}tre} G} (\bibinfo{year}{1927}), \bibinfo{month}{Jan.}
\bibinfo{title}{{Un Univers homog{\`e}ne de masse constante et de rayon croissant rendant compte de la vitesse radiale des n{\'e}buleuses extra-galactiques}}.
\bibinfo{journal}{{\em Annales de la Soci{\'e}t{\'e} Scientifique de Bruxelles}} \bibinfo{volume}{47}: \bibinfo{pages}{49--59}.

\bibtype{Article}%
\bibitem[{Linder}(2003)]{Linder2003}
\bibinfo{author}{{Linder} EV} (\bibinfo{year}{2003}), \bibinfo{month}{Mar.}
\bibinfo{title}{{Exploring the Expansion History of the Universe}}.
\bibinfo{journal}{{\em \prl}} \bibinfo{volume}{90} (\bibinfo{number}{9}), \bibinfo{eid}{091301}. \bibinfo{doi}{\doi{10.1103/PhysRevLett.90.091301}}.
\eprint{astro-ph/0208512}.

\bibtype{Book}%
\bibitem[{Moore}(2013)]{Moore2013}
\bibinfo{author}{{Moore} TA} (\bibinfo{year}{2013}).
\bibinfo{title}{{A General Relativity Workbook}}, \bibinfo{address}{University Science Books}.

\bibtype{Article}%
\bibitem[{Ntelis} et al.(2017)]{Ntelis2017}
\bibinfo{author}{{Ntelis} P}, \bibinfo{author}{{Hamilton} JC}, \bibinfo{author}{{Le Goff} JM}, \bibinfo{author}{{Burtin} E}, \bibinfo{author}{{Laurent} P}, \bibinfo{author}{{Rich} J}, \bibinfo{author}{{Guillermo Busca} N}, \bibinfo{author}{{Tinker} J}, \bibinfo{author}{{Aubourg} E}, \bibinfo{author}{{du Mas des Bourboux} H}, \bibinfo{author}{{Bautista} J}, \bibinfo{author}{{Palanque Delabrouille} N}, \bibinfo{author}{{Delubac} T}, \bibinfo{author}{{Eftekharzadeh} S}, \bibinfo{author}{{Hogg} DW}, \bibinfo{author}{{Myers} A}, \bibinfo{author}{{Vargas-Maga{\~n}a} M}, \bibinfo{author}{{P{\^a}ris} I}, \bibinfo{author}{{Petitjean} P}, \bibinfo{author}{{Rossi} G}, \bibinfo{author}{{Schneider} DP}, \bibinfo{author}{{Tojeiro} R} and  \bibinfo{author}{{Yeche} C} (\bibinfo{year}{2017}), \bibinfo{month}{Jun.}
\bibinfo{title}{{Exploring cosmic homogeneity with the BOSS DR12 galaxy sample}}.
\bibinfo{journal}{{\em \jcap}} \bibinfo{volume}{2017} (\bibinfo{number}{6}), \bibinfo{eid}{019}. \bibinfo{doi}{\doi{10.1088/1475-7516/2017/06/019}}.
\eprint{1702.02159}.

\bibtype{Article}%
\bibitem[{Perlmutter} et al.(1999)]{Perlmutter1999}
\bibinfo{author}{{Perlmutter} S}, \bibinfo{author}{{Aldering} G}, \bibinfo{author}{{Goldhaber} G}, \bibinfo{author}{{Knop} RA}, \bibinfo{author}{{Nugent} P}, \bibinfo{author}{{Castro} PG}, \bibinfo{author}{{Deustua} S}, \bibinfo{author}{{Fabbro} S}, \bibinfo{author}{{Goobar} A}, \bibinfo{author}{{Groom} DE}, \bibinfo{author}{{Hook} IM}, \bibinfo{author}{{Kim} AG}, \bibinfo{author}{{Kim} MY}, \bibinfo{author}{{Lee} JC}, \bibinfo{author}{{Nunes} NJ}, \bibinfo{author}{{Pain} R}, \bibinfo{author}{{Pennypacker} CR}, \bibinfo{author}{{Quimby} R}, \bibinfo{author}{{Lidman} C}, \bibinfo{author}{{Ellis} RS}, \bibinfo{author}{{Irwin} M}, \bibinfo{author}{{McMahon} RG}, \bibinfo{author}{{Ruiz-Lapuente} P}, \bibinfo{author}{{Walton} N}, \bibinfo{author}{{Schaefer} B}, \bibinfo{author}{{Boyle} BJ}, \bibinfo{author}{{Filippenko} AV}, \bibinfo{author}{{Matheson} T}, \bibinfo{author}{{Fruchter} AS}, \bibinfo{author}{{Panagia} N}, \bibinfo{author}{{Newberg} HJM}, \bibinfo{author}{{Couch} WJ} and  \bibinfo{author}{{Project}
  TSC} (\bibinfo{year}{1999}), \bibinfo{month}{Jun.}
\bibinfo{title}{{Measurements of {\ensuremath{\Omega}} and {\ensuremath{\Lambda}} from 42 High-Redshift Supernovae}}.
\bibinfo{journal}{{\em \apj}} \bibinfo{volume}{517} (\bibinfo{number}{2}): \bibinfo{pages}{565--586}. \bibinfo{doi}{\doi{10.1086/307221}}.
\eprint{astro-ph/9812133}.

\bibtype{Article}%
\bibitem[{Planck Collaboration} et al.(2020)]{Planck18_VI}
\bibinfo{author}{{Planck Collaboration}}, \bibinfo{author}{{Aghanim} N}, \bibinfo{author}{{Akrami} Y}, \bibinfo{author}{{Ashdown} M}, \bibinfo{author}{{Aumont} J}, \bibinfo{author}{{Baccigalupi} C}, \bibinfo{author}{{Ballardini} M}, \bibinfo{author}{{Banday} AJ}, \bibinfo{author}{{Barreiro} RB}, \bibinfo{author}{{Bartolo} N}, \bibinfo{author}{{Basak} S}, \bibinfo{author}{{Battye} R}, \bibinfo{author}{{Benabed} K}, \bibinfo{author}{{Bernard} JP}, \bibinfo{author}{{Bersanelli} M}, \bibinfo{author}{{Bielewicz} P}, \bibinfo{author}{{Bock} JJ}, \bibinfo{author}{{Bond} JR}, \bibinfo{author}{{Borrill} J}, \bibinfo{author}{{Bouchet} FR}, \bibinfo{author}{{Boulanger} F}, \bibinfo{author}{{Bucher} M}, \bibinfo{author}{{Burigana} C}, \bibinfo{author}{{Butler} RC}, \bibinfo{author}{{Calabrese} E}, \bibinfo{author}{{Cardoso} JF}, \bibinfo{author}{{Carron} J}, \bibinfo{author}{{Challinor} A}, \bibinfo{author}{{Chiang} HC}, \bibinfo{author}{{Chluba} J}, \bibinfo{author}{{Colombo} LPL}, \bibinfo{author}{{Combet} C},
  \bibinfo{author}{{Contreras} D}, \bibinfo{author}{{Crill} BP}, \bibinfo{author}{{Cuttaia} F}, \bibinfo{author}{{de Bernardis} P}, \bibinfo{author}{{de Zotti} G}, \bibinfo{author}{{Delabrouille} J}, \bibinfo{author}{{Delouis} JM}, \bibinfo{author}{{Di Valentino} E}, \bibinfo{author}{{Diego} JM}, \bibinfo{author}{{Dor{\'e}} O}, \bibinfo{author}{{Douspis} M}, \bibinfo{author}{{Ducout} A}, \bibinfo{author}{{Dupac} X}, \bibinfo{author}{{Dusini} S}, \bibinfo{author}{{Efstathiou} G}, \bibinfo{author}{{Elsner} F}, \bibinfo{author}{{En{\ss}lin} TA}, \bibinfo{author}{{Eriksen} HK}, \bibinfo{author}{{Fantaye} Y}, \bibinfo{author}{{Farhang} M}, \bibinfo{author}{{Fergusson} J}, \bibinfo{author}{{Fernandez-Cobos} R}, \bibinfo{author}{{Finelli} F}, \bibinfo{author}{{Forastieri} F}, \bibinfo{author}{{Frailis} M}, \bibinfo{author}{{Fraisse} AA}, \bibinfo{author}{{Franceschi} E}, \bibinfo{author}{{Frolov} A}, \bibinfo{author}{{Galeotta} S}, \bibinfo{author}{{Galli} S}, \bibinfo{author}{{Ganga} K},
  \bibinfo{author}{{G{\'e}nova-Santos} RT}, \bibinfo{author}{{Gerbino} M}, \bibinfo{author}{{Ghosh} T}, \bibinfo{author}{{Gonz{\'a}lez-Nuevo} J}, \bibinfo{author}{{G{\'o}rski} KM}, \bibinfo{author}{{Gratton} S}, \bibinfo{author}{{Gruppuso} A}, \bibinfo{author}{{Gudmundsson} JE}, \bibinfo{author}{{Hamann} J}, \bibinfo{author}{{Handley} W}, \bibinfo{author}{{Hansen} FK}, \bibinfo{author}{{Herranz} D}, \bibinfo{author}{{Hildebrandt} SR}, \bibinfo{author}{{Hivon} E}, \bibinfo{author}{{Huang} Z}, \bibinfo{author}{{Jaffe} AH}, \bibinfo{author}{{Jones} WC}, \bibinfo{author}{{Karakci} A}, \bibinfo{author}{{Keih{\"a}nen} E}, \bibinfo{author}{{Keskitalo} R}, \bibinfo{author}{{Kiiveri} K}, \bibinfo{author}{{Kim} J}, \bibinfo{author}{{Kisner} TS}, \bibinfo{author}{{Knox} L}, \bibinfo{author}{{Krachmalnicoff} N}, \bibinfo{author}{{Kunz} M}, \bibinfo{author}{{Kurki-Suonio} H}, \bibinfo{author}{{Lagache} G}, \bibinfo{author}{{Lamarre} JM}, \bibinfo{author}{{Lasenby} A}, \bibinfo{author}{{Lattanzi} M},
  \bibinfo{author}{{Lawrence} CR}, \bibinfo{author}{{Le Jeune} M}, \bibinfo{author}{{Lemos} P}, \bibinfo{author}{{Lesgourgues} J}, \bibinfo{author}{{Levrier} F}, \bibinfo{author}{{Lewis} A}, \bibinfo{author}{{Liguori} M}, \bibinfo{author}{{Lilje} PB}, \bibinfo{author}{{Lilley} M}, \bibinfo{author}{{Lindholm} V}, \bibinfo{author}{{L{\'o}pez-Caniego} M}, \bibinfo{author}{{Lubin} PM}, \bibinfo{author}{{Ma} YZ}, \bibinfo{author}{{Mac{\'\i}as-P{\'e}rez} JF}, \bibinfo{author}{{Maggio} G}, \bibinfo{author}{{Maino} D}, \bibinfo{author}{{Mandolesi} N}, \bibinfo{author}{{Mangilli} A}, \bibinfo{author}{{Marcos-Caballero} A}, \bibinfo{author}{{Maris} M}, \bibinfo{author}{{Martin} PG}, \bibinfo{author}{{Martinelli} M}, \bibinfo{author}{{Mart{\'\i}nez-Gonz{\'a}lez} E}, \bibinfo{author}{{Matarrese} S}, \bibinfo{author}{{Mauri} N}, \bibinfo{author}{{McEwen} JD}, \bibinfo{author}{{Meinhold} PR}, \bibinfo{author}{{Melchiorri} A}, \bibinfo{author}{{Mennella} A}, \bibinfo{author}{{Migliaccio} M}, \bibinfo{author}{{Millea} M},
  \bibinfo{author}{{Mitra} S}, \bibinfo{author}{{Miville-Desch{\^e}nes} MA}, \bibinfo{author}{{Molinari} D}, \bibinfo{author}{{Montier} L}, \bibinfo{author}{{Morgante} G}, \bibinfo{author}{{Moss} A}, \bibinfo{author}{{Natoli} P}, \bibinfo{author}{{N{\o}rgaard-Nielsen} HU}, \bibinfo{author}{{Pagano} L}, \bibinfo{author}{{Paoletti} D}, \bibinfo{author}{{Partridge} B}, \bibinfo{author}{{Patanchon} G}, \bibinfo{author}{{Peiris} HV}, \bibinfo{author}{{Perrotta} F}, \bibinfo{author}{{Pettorino} V}, \bibinfo{author}{{Piacentini} F}, \bibinfo{author}{{Polastri} L}, \bibinfo{author}{{Polenta} G}, \bibinfo{author}{{Puget} JL}, \bibinfo{author}{{Rachen} JP}, \bibinfo{author}{{Reinecke} M}, \bibinfo{author}{{Remazeilles} M}, \bibinfo{author}{{Renzi} A}, \bibinfo{author}{{Rocha} G}, \bibinfo{author}{{Rosset} C}, \bibinfo{author}{{Roudier} G}, \bibinfo{author}{{Rubi{\~n}o-Mart{\'\i}n} JA}, \bibinfo{author}{{Ruiz-Granados} B}, \bibinfo{author}{{Salvati} L}, \bibinfo{author}{{Sandri} M}, \bibinfo{author}{{Savelainen} M},
  \bibinfo{author}{{Scott} D}, \bibinfo{author}{{Shellard} EPS}, \bibinfo{author}{{Sirignano} C}, \bibinfo{author}{{Sirri} G}, \bibinfo{author}{{Spencer} LD}, \bibinfo{author}{{Sunyaev} R}, \bibinfo{author}{{Suur-Uski} AS}, \bibinfo{author}{{Tauber} JA}, \bibinfo{author}{{Tavagnacco} D}, \bibinfo{author}{{Tenti} M}, \bibinfo{author}{{Toffolatti} L}, \bibinfo{author}{{Tomasi} M}, \bibinfo{author}{{Trombetti} T}, \bibinfo{author}{{Valenziano} L}, \bibinfo{author}{{Valiviita} J}, \bibinfo{author}{{Van Tent} B}, \bibinfo{author}{{Vibert} L}, \bibinfo{author}{{Vielva} P}, \bibinfo{author}{{Villa} F}, \bibinfo{author}{{Vittorio} N}, \bibinfo{author}{{Wand elt} BD}, \bibinfo{author}{{Wehus} IK}, \bibinfo{author}{{White} M}, \bibinfo{author}{{White} SDM}, \bibinfo{author}{{Zacchei} A} and  \bibinfo{author}{{Zonca} A} (\bibinfo{year}{2020}), \bibinfo{month}{Sep.}
\bibinfo{title}{{Planck 2018 results. VI. Cosmological parameters}}.
\bibinfo{journal}{{\em \aap}} \bibinfo{volume}{641}, \bibinfo{eid}{A6}. \bibinfo{doi}{\doi{10.1051/0004-6361/201833910}}.
\eprint{1807.06209}.

\bibtype{Article}%
\bibitem[{Riess} et al.(1998)]{Riess1998}
\bibinfo{author}{{Riess} AG}, \bibinfo{author}{{Filippenko} AV}, \bibinfo{author}{{Challis} P}, \bibinfo{author}{{Clocchiatti} A}, \bibinfo{author}{{Diercks} A}, \bibinfo{author}{{Garnavich} PM}, \bibinfo{author}{{Gilliland} RL}, \bibinfo{author}{{Hogan} CJ}, \bibinfo{author}{{Jha} S}, \bibinfo{author}{{Kirshner} RP}, \bibinfo{author}{{Leibundgut} B}, \bibinfo{author}{{Phillips} MM}, \bibinfo{author}{{Reiss} D}, \bibinfo{author}{{Schmidt} BP}, \bibinfo{author}{{Schommer} RA}, \bibinfo{author}{{Smith} RC}, \bibinfo{author}{{Spyromilio} J}, \bibinfo{author}{{Stubbs} C}, \bibinfo{author}{{Suntzeff} NB} and  \bibinfo{author}{{Tonry} J} (\bibinfo{year}{1998}), \bibinfo{month}{Sep.}
\bibinfo{title}{{Observational Evidence from Supernovae for an Accelerating Universe and a Cosmological Constant}}.
\bibinfo{journal}{{\em \aj}} \bibinfo{volume}{116} (\bibinfo{number}{3}): \bibinfo{pages}{1009--1038}. \bibinfo{doi}{\doi{10.1086/300499}}.
\eprint{astro-ph/9805201}.

\bibtype{Article}%
\bibitem[{Riess} et al.(2022)]{Riess2022}
\bibinfo{author}{{Riess} AG}, \bibinfo{author}{{Yuan} W}, \bibinfo{author}{{Macri} LM}, \bibinfo{author}{{Scolnic} D}, \bibinfo{author}{{Brout} D}, \bibinfo{author}{{Casertano} S}, \bibinfo{author}{{Jones} DO}, \bibinfo{author}{{Murakami} Y}, \bibinfo{author}{{Anand} GS}, \bibinfo{author}{{Breuval} L}, \bibinfo{author}{{Brink} TG}, \bibinfo{author}{{Filippenko} AV}, \bibinfo{author}{{Hoffmann} S}, \bibinfo{author}{{Jha} SW}, \bibinfo{author}{{D'arcy Kenworthy} W}, \bibinfo{author}{{Mackenty} J}, \bibinfo{author}{{Stahl} BE} and  \bibinfo{author}{{Zheng} W} (\bibinfo{year}{2022}), \bibinfo{month}{Jul.}
\bibinfo{title}{{A Comprehensive Measurement of the Local Value of the Hubble Constant with 1 km s$^{-1}$ Mpc$^{-1}$ Uncertainty from the Hubble Space Telescope and the SH0ES Team}}.
\bibinfo{journal}{{\em \apjl}} \bibinfo{volume}{934} (\bibinfo{number}{1}), \bibinfo{eid}{L7}. \bibinfo{doi}{\doi{10.3847/2041-8213/ac5c5b}}.
\eprint{2112.04510}.

\bibtype{Article}%
\bibitem[{Scrimgeour} et al.(2012)]{Scrimgeour2012}
\bibinfo{author}{{Scrimgeour} MI}, \bibinfo{author}{{Davis} T}, \bibinfo{author}{{Blake} C}, \bibinfo{author}{{James} JB}, \bibinfo{author}{{Poole} GB}, \bibinfo{author}{{Staveley-Smith} L}, \bibinfo{author}{{Brough} S}, \bibinfo{author}{{Colless} M}, \bibinfo{author}{{Contreras} C}, \bibinfo{author}{{Couch} W}, \bibinfo{author}{{Croom} S}, \bibinfo{author}{{Croton} D}, \bibinfo{author}{{Drinkwater} MJ}, \bibinfo{author}{{Forster} K}, \bibinfo{author}{{Gilbank} D}, \bibinfo{author}{{Gladders} M}, \bibinfo{author}{{Glazebrook} K}, \bibinfo{author}{{Jelliffe} B}, \bibinfo{author}{{Jurek} RJ}, \bibinfo{author}{{Li} Ih}, \bibinfo{author}{{Madore} B}, \bibinfo{author}{{Martin} DC}, \bibinfo{author}{{Pimbblet} K}, \bibinfo{author}{{Pracy} M}, \bibinfo{author}{{Sharp} R}, \bibinfo{author}{{Wisnioski} E}, \bibinfo{author}{{Woods} D}, \bibinfo{author}{{Wyder} TK} and  \bibinfo{author}{{Yee} HKC} (\bibinfo{year}{2012}), \bibinfo{month}{Sep.}
\bibinfo{title}{{The WiggleZ Dark Energy Survey: the transition to large-scale cosmic homogeneity}}.
\bibinfo{journal}{{\em \mnras}} \bibinfo{volume}{425} (\bibinfo{number}{1}): \bibinfo{pages}{116--134}. \bibinfo{doi}{\doi{10.1111/j.1365-2966.2012.21402.x}}.
\eprint{1205.6812}.

\bibtype{Article}%
\bibitem[{Watkins} et al.(2023)]{Watkins2023}
\bibinfo{author}{{Watkins} R}, \bibinfo{author}{{Allen} T}, \bibinfo{author}{{Bradford} CJ}, \bibinfo{author}{{Ramon} A}, \bibinfo{author}{{Walker} A}, \bibinfo{author}{{Feldman} HA}, \bibinfo{author}{{Cionitti} R}, \bibinfo{author}{{Al-Shorman} Y}, \bibinfo{author}{{Kourkchi} E} and  \bibinfo{author}{{Tully} RB} (\bibinfo{year}{2023}), \bibinfo{month}{Sep.}
\bibinfo{title}{{Analysing the large-scale bulk flow using cosmicflows4: increasing tension with the standard cosmological model}}.
\bibinfo{journal}{{\em \mnras}} \bibinfo{volume}{524} (\bibinfo{number}{2}): \bibinfo{pages}{1885--1892}. \bibinfo{doi}{\doi{10.1093/mnras/stad1984}}.
\eprint{2302.02028}.

\bibtype{Article}%
\bibitem[{Whitford} et al.(2023)]{Whitford2023}
\bibinfo{author}{{Whitford} AM}, \bibinfo{author}{{Howlett} C} and  \bibinfo{author}{{Davis} TM} (\bibinfo{year}{2023}), \bibinfo{month}{Dec.}
\bibinfo{title}{{Evaluating bulk flow estimators for CosmicFlows-4 measurements}}.
\bibinfo{journal}{{\em \mnras}} \bibinfo{volume}{526} (\bibinfo{number}{2}): \bibinfo{pages}{3051--3071}. \bibinfo{doi}{\doi{10.1093/mnras/stad2764}}.
\eprint{2306.11269}.

\bibtype{Article}%
\bibitem[{Wiltshire}(2007)]{Wiltshire2007}
\bibinfo{author}{{Wiltshire} DL} (\bibinfo{year}{2007}), \bibinfo{month}{Dec.}
\bibinfo{title}{{Exact Solution to the Averaging Problem in Cosmology}}.
\bibinfo{journal}{{\em \prl}} \bibinfo{volume}{99} (\bibinfo{number}{25}), \bibinfo{eid}{251101}. \bibinfo{doi}{\doi{10.1103/PhysRevLett.99.251101}}.
\eprint{0709.0732}.

\bibtype{Article}%
\bibitem[{Wright} et al.(2025)]{KIDS2025}
\bibinfo{author}{{Wright} AH}, \bibinfo{author}{{St{\"o}lzner} B}, \bibinfo{author}{{Asgari} M}, \bibinfo{author}{{Bilicki} M}, \bibinfo{author}{{Giblin} B}, \bibinfo{author}{{Heymans} C}, \bibinfo{author}{{Hildebrandt} H}, \bibinfo{author}{{Hoekstra} H}, \bibinfo{author}{{Joachimi} B}, \bibinfo{author}{{Kuijken} K}, \bibinfo{author}{{Li} SS}, \bibinfo{author}{{Reischke} R}, \bibinfo{author}{{von Wietersheim-Kramsta} M}, \bibinfo{author}{{Yoon} M}, \bibinfo{author}{{Burger} P}, \bibinfo{author}{{Chisari} NE}, \bibinfo{author}{{de Jong} J}, \bibinfo{author}{{Dvornik} A}, \bibinfo{author}{{Georgiou} C}, \bibinfo{author}{{Harnois-D{\'e}raps} J}, \bibinfo{author}{{Jalan} P}, \bibinfo{author}{{William} AJ}, \bibinfo{author}{{Joudaki} S}, \bibinfo{author}{{Lesci} GF}, \bibinfo{author}{{Linke} L}, \bibinfo{author}{{Loureiro} A}, \bibinfo{author}{{Mahony} C}, \bibinfo{author}{{Maturi} M}, \bibinfo{author}{{Miller} L}, \bibinfo{author}{{Moscardini} L}, \bibinfo{author}{{Napolitano} NR}, \bibinfo{author}{{Porth} L},
  \bibinfo{author}{{Radovich} M}, \bibinfo{author}{{Schneider} P}, \bibinfo{author}{{Tr{\"o}ster} T}, \bibinfo{author}{{Wittje} A}, \bibinfo{author}{{Yan} Z} and  \bibinfo{author}{{Zhang} YH} (\bibinfo{year}{2025}), \bibinfo{month}{Mar.}
\bibinfo{title}{{KiDS-Legacy: Cosmological constraints from cosmic shear with the complete Kilo-Degree Survey}}.
\bibinfo{journal}{{\em arXiv e-prints}} , \bibinfo{eid}{arXiv:2503.19441}\bibinfo{doi}{\doi{10.48550/arXiv.2503.19441}}.
\eprint{2503.19441}.

\end{thebibliography*}

\end{document}